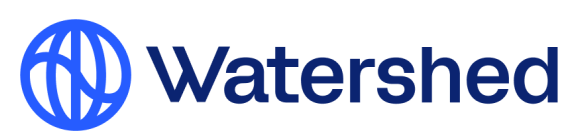

# Estimating GHG Emissions from AI Use

## Framework for Corporate-Level Measurement

## Summary

Electricity demand from data centers is expected to grow from roughly 5% of U.S. consumption in 2025 to between 9% and 17% by 2030, and corporate artificial intelligence (AI) use is following a similar trajectory, spanning employee productivity assistants, direct access to large language models (LLMs), and AI features embedded in enterprise software. AI emissions today are a small share of footprints for many enterprises, but that share is unlikely to remain small for long. Without reasonable estimates, companies cannot set reduction targets or identify effective decarbonization levers as emissions grow. Companies, regulators, and auditors are asking for emissions estimates that withstand scrutiny, but no widely accepted methodology exists today. Published per-query estimates can differ by several orders of magnitude depending on what is counted, which provider is measured, and what assumptions are made about electricity use and the grid mix.

This white paper proposes a standardized framework for corporate-level AI emissions accounting. The framework is designed to be **defensible with current data constraints, tiered to meet companies where their data are, transparent about its assumptions, updatable as provider disclosure matures, and built for action rather than disclosure alone.** Since AI emissions accounting is still nascent, it has the opportunity to design for actionability from the outset, so that measurement incentivizes responsible choices during AI's rapid buildout.

**The framework has three elements:**

1. The *system boundary* is designed to be comprehensive, including inference compute, training, data center overhead, host system capacity, idle and reserved capacity, and amortized embodied emissions of accelerator hardware and supporting infrastructure.
2. The *functional unit* is kilograms of $CO_2$e per million tokens for inference, reported alongside electricity consumption so that customers can consider strategies for reducing emissions and matching clean electricity. Tokens are also the metric that many companies track for cost reasons, which allows token-level emissions reporting to leverage engineering efforts.
3. The *three-tier calculation approach* runs from the Spend Tier (a spend-based backstop) to the Activity Tier (activity-based estimates using token volumes and modeled energy intensity) to the Provider Tier (provider-reported per-token figures, which is the most precise tier).[1] Electricity consumption per task varies widely across workload types, from chat prompts to reasoning and agentic workflows (Figure ES-1).

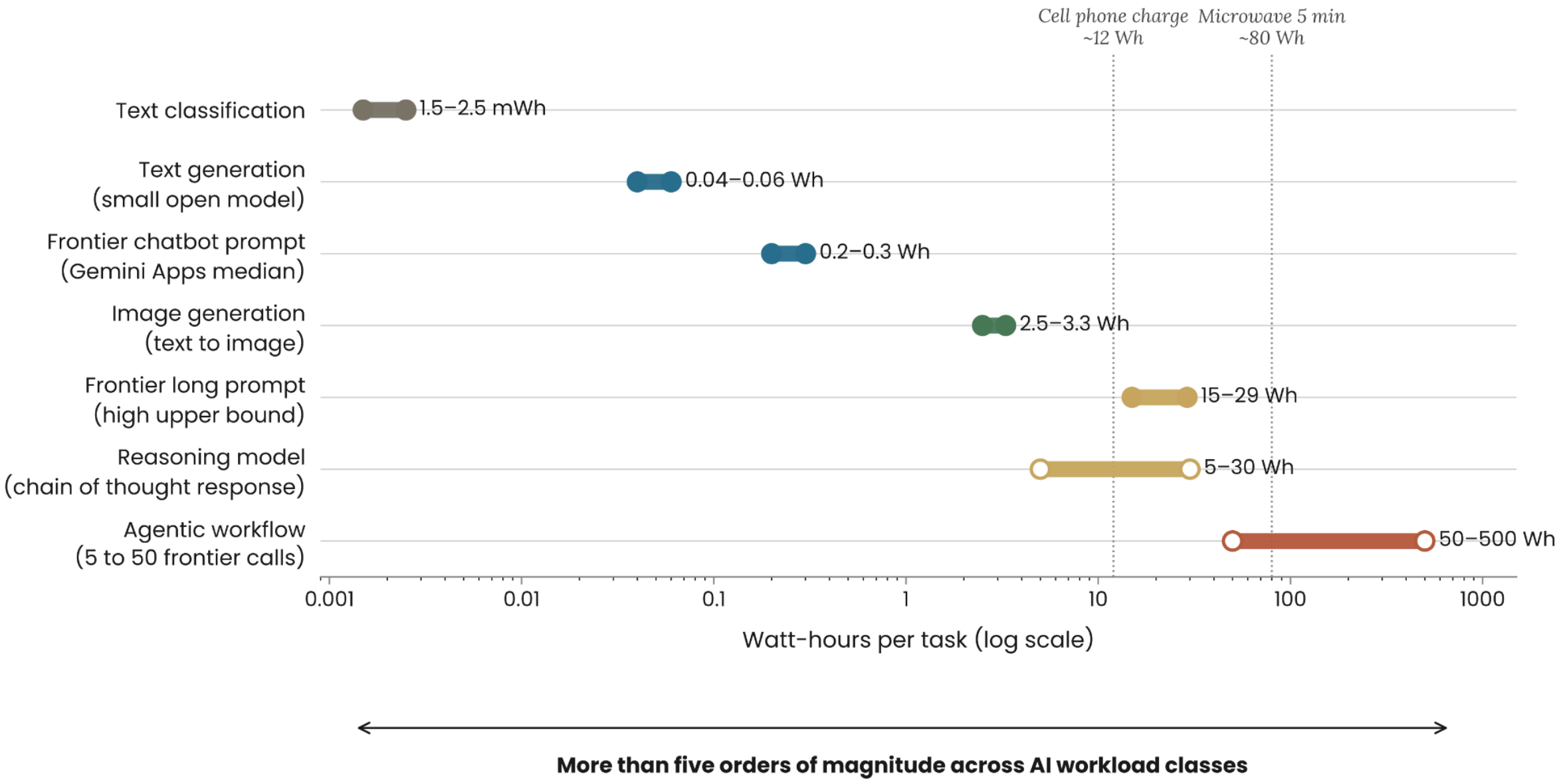

**Figure ES-1: Electricity consumption per AI task across different workloads.** Electricity use per task varies by more than two orders of magnitude across workload types. The closed circles in ranges represent values that come directly from the literature. The open circles represent values derived from per-prompt energy intensities. See Figure 2 for a detailed list of included studies and models.

To illustrate the stakes, a stylized example shows a four-fold range of estimates for a company with $100,000 annual spending on a frontier model, accessed programmatically for internal productivity tools and a customer-facing chatbot. The same AI use yields **13.4 t$CO_2$e under the Spend Tier, 3.7-5.4 t$CO_2$e under the Activity Tier, and 3.2-4.5 t$CO_2$e under the Provider Tier.** Spend-based methods can overstate or understate emissions depending on the context. In this illustrative example, the company's premium-priced frontier API implies less electricity per dollar than the sector average, producing an overestimate, but this direction could reverse if, for instance, providers subsidize usage or are operating on more carbon-intensive grids. Training emissions contribute between 32% and 56% of the footprint depending on assumptions about total lifetime token volume, which is a parameter no frontier model currently discloses. For widely deployed models, however, the more actionable data gap is inference. **Operational electricity is likely a larger share of total emissions across the**

[1] Reaching the most detailed accounting tiers requires provider disclosures that largely do not yet exist, especially for frontier models, a gap discussed later. Even where per-token figures are published, they tend to be fleet-wide estimates for a typical prompt rather than values linked to specific customer models, regions, or traffic.

**lifetime of heavily used models and is the component most directly addressable through clean electricity procurement, region selection, and provider efficiency improvements.**

The accuracy of estimates depends on parameters from model providers and their cloud hosts. The framework suggests a standardized disclosure table to close gaps for frontier models, which includes per-model electricity intensities per token, grid emissions factors, embodied and training emissions, and data center overhead.

Data transparency may also be in providers' commercial self-interest: when better data are available, **AI's footprint has typically been materially lower than published estimates suggest, as non-production benchmarks can overstate inference energy by 4 to 20 times by omitting efficiency gains from deployment at scale,** including batching, caching, and hardware co-design optimizations that are proprietary and therefore absent from external estimates.[2] A provider that discloses figures lowers emissions its customers report and prevents its products from being misrepresented by third-party estimates. Moving the field toward the provider-reported tier has data benefits for providers and customers alike.

For customers, the framework suggests a set of practical management levers that operate through two channels. Prompt design and model selection reduce electricity consumed per task (**reasoning models can be 30 times more energy intensive** than smaller variants), while region selection and clean energy matching reduce the carbon intensity of each kWh consumed (**U.S. subregions vary by more than five-fold in their carbon intensity**). Reporting electricity alongside emissions keeps both channels visible and lets companies attribute year-over-year changes to efficiency gains rather than grid decarbonization or vice versa.

The framework is designed to establish a methodology, data flows, and provider relationships that will be valuable as AI emissions grow and to **demystify AI's footprint with auditable data before inflated narratives harden into convention.** The emphasis throughout is on accounting that drives action to reduce emissions – efficiency, model choice, region, clean electricity matching – in addition to reporting them.

**Getting Started in Four Steps**

1. *Inventory your AI use by channel.* List vendors and how each is accessed (e.g., hosted assistants, embedded features) and pull AI-related data.
2. *Estimate at the tier your data supports.* Use the Spend Tier as a backstop for everything, and move channels where you have token volumes to the Activity Tier.
3. *Ask providers for better data.* Request the essential fields in Table 5, starting with per-token electricity intensity and serving region, and upgrade to the Provider Tier

[2] Oviedo, Felipe, et al. "Energy Use of AI Inference, Efficiency Pathways, and Test-Time Scaling." *Joule* (2026). This study focuses on open-weight models.

where disclosed.
4. *Report both metrics and act on the levers.* Report emissions alongside electricity, then apply the management levers: model routing, prompt design, region selection, and clean electricity matching.

# Introduction

Corporate sustainability teams, investors, and regulators are increasingly asking about how to account for greenhouse gas (GHG) emissions from AI use by employees and enterprise tools and finding that no widely accepted methodology exists. These questions reflect burgeoning AI and data center demand, which is expected to consume 9% to 17% of U.S. electricity by 2030, up from ~5% in 2025.[3]

As later sections discuss, estimates in the literature have inconsistent system boundaries, varying energy intensity factors, different grid emissions methods, and omit embodied emissions, idle capacity, and other factors that materially alter environmental impacts.

Several dynamic trends are driving AI emissions in opposing directions. Algorithmic efficiency for a given level of performance has roughly halved every eight months (Ho, et al., 2024),[4] and inference-side optimizations (e.g., batching, KV and prompt caching, speculative decoding, quantized inference) and optimized software and hardware co-design can dramatically reduce per-query energy. At the same time, **AI use is shifting toward higher per-task compute intensity with reasoning models generating extended inference; multi-modal systems handling text, image, audio, and video; and agentic systems executing multi-step workflows autonomously**, which means that a single request triggers many model calls.

This paper aims to help stakeholders understand and address AI-related emissions and electricity demand using a **standardized methodology that is designed to be defensible, transparent, and updatable.** The objectives of this paper are to:

- Establish a framework for estimating the electricity consumption and GHG footprint of corporate AI use, including empirically grounded data and broad system boundaries to accurately reflect environmental impacts, encourage data gathering, and incentivize actions to reduce emissions.

[3] EPRI (2026), "Powering Intelligence," EPRI Report 3002034696. Available at: https://powering-intelligence.epri.com/. Data centers serve a wide variety of workloads beyond AI, including streaming, cloud storage, and enterprise software. AI represents a growing share of total data center electricity, hence the focus of this framework.
[4] Ho, Anson, et al. "Algorithmic Progress in Language Models." *Advances in Neural Information Processing Systems* 37 (2024).

- Apply this framework to produce illustrative estimates using the best-available public information.
- Highlight key data gaps for customers calculating their Scope 3 footprint from AI services. Rapid changes to models, use cases, and grids mean that the framework should be updated frequently.
- Design the accounting to support action. The framework treats disclosure and decision-usefulness as complementary and reports electricity alongside emissions, specifically so that measurement points toward concrete levers for reducing impact.

# Why Is Measuring AI Emissions Hard?

1. Limited Data from First-Party Providers

Accurate AI emissions accounting[5] is constrained by the tension between the need for transparency and legitimate commercial sensitivity of providers. Many parameters that are most valuable for emissions accounting (e.g., model-specific inference energy, training energy, batch sizes, hardware utilization) overlap with information that providers have reasonable business interests in protecting. This is especially true of efficiency gains: because the most recent hardware configurations, software optimizations, and serving infrastructure improvements are proprietary, external estimates based on publicly available benchmarks tend to lag actual performance.[6]

This dynamic suggests that **emissions estimates from public data can considerably overstate the footprint of frontier models.** Oviedo, et al. (2026) find that **benchmarks overstate inference energy by 4 to 20 times** by failing to account for efficiency gains that are only realized at production scale, including batching, KV caching, speculative decoding, and optimized hardware co-design. A standardized disclosure framework could narrow this gap, even one that protects the most commercially sensitive details.

A related data challenge is the quickly changing landscapes of AI model architectures and user behavior (Figure 1). Alternative assumptions (e.g., data center efficiency, hardware

---

[5] Throughout this paper, "AI emissions" primarily refers to GHG emissions attributable to a company's use of third-party AI services, which includes employee productivity assistants, direct access to LLMs, and AI features embedded in enterprise software. Contracted AI use is primarily classified as Scope 3, Category 1 (Purchased Goods and Services) for many companies, whether that service is accessed directly from the model provider or through a cloud intermediary. Emissions from training or self-hosting a company's own models are accounted for directly through that company's own compute and are outside this paper's scope.
[6] Masanet, Eric, et al. "To Better Understand AI's Growing Energy Use, Analysts Need a Data Revolution." *Joule* 8 (2024).

choice, grid carbon intensity) create orders-of-magnitude variation in emissions for equivalent workloads.[7]

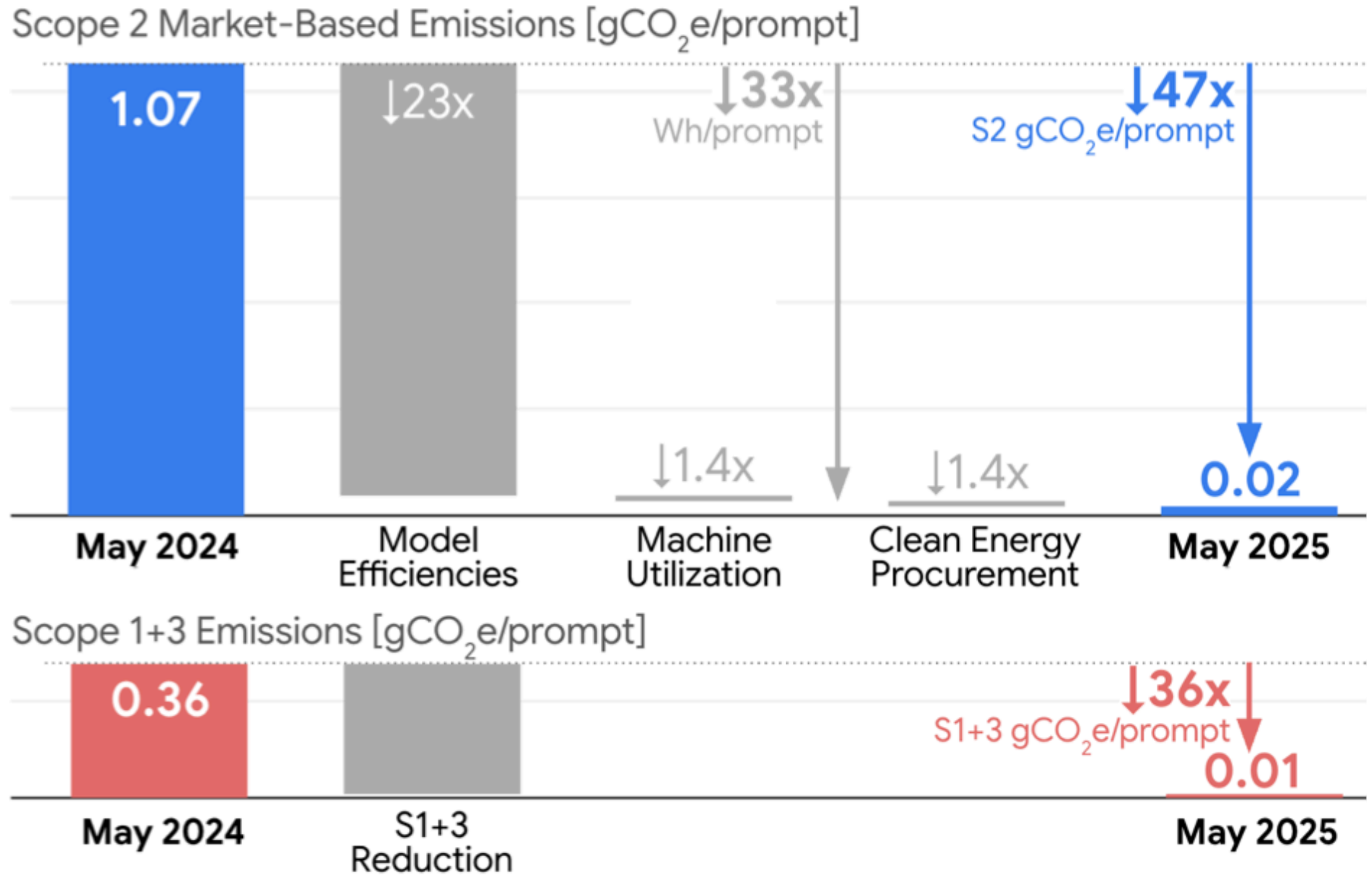

**Figure 1: Median Gemini Apps text prompt emissions over time.** Footprint is decomposed into Scope 2 market-based emissions (top) and Scopes 1 and 3 emissions (bottom). Google reports a 47-fold reduction in the energy cost of a Gemini prompt over a single year. Source: Elsworth, et al. (2025).[8]

## 2. System Boundary Ambiguity and Disagreement

Typical LLM queries run through many layers with different energy and emissions impacts. Published per-query numbers can differ by several orders of magnitude depending on what is counted in terms of embodied hardware, training, and inference. In addition, there is no agreed-upon functional unit for emissions comparison. Generative AI outputs (e.g., per query, per token, per image) vary widely in their computational intensity.

## 3. Heterogeneity in AI Use Cases

Emissions estimates are further complicated by the heterogeneity of enterprise AI use: prompt complexity, modality, token length, and degree of agentic or reasoning-intensive computation all vary substantially within and across companies. Each dimension can shift energy consumption by orders of magnitude relative to a simple chat query (Figure 2).[9]

[7] Efficiency improvements also can make technologies cheaper and more accessible, driving greater adoption, which is known as the Jevons paradox. AI's low per-query cost unlocks new use cases, including multi-agent workflows for customer support, automated image and video generation, and coding assistance. This effect means that, even if a simple query has a similar energy intensity to a web-based search, total consumption can increase many-fold due to a proliferation of use cases that were not economically viable.
[8] Elsworth, Cooper, et al. "Measuring the Environmental Impact of Delivering AI at Google Scale." arXiv preprint arXiv:2508.15734 (2025).

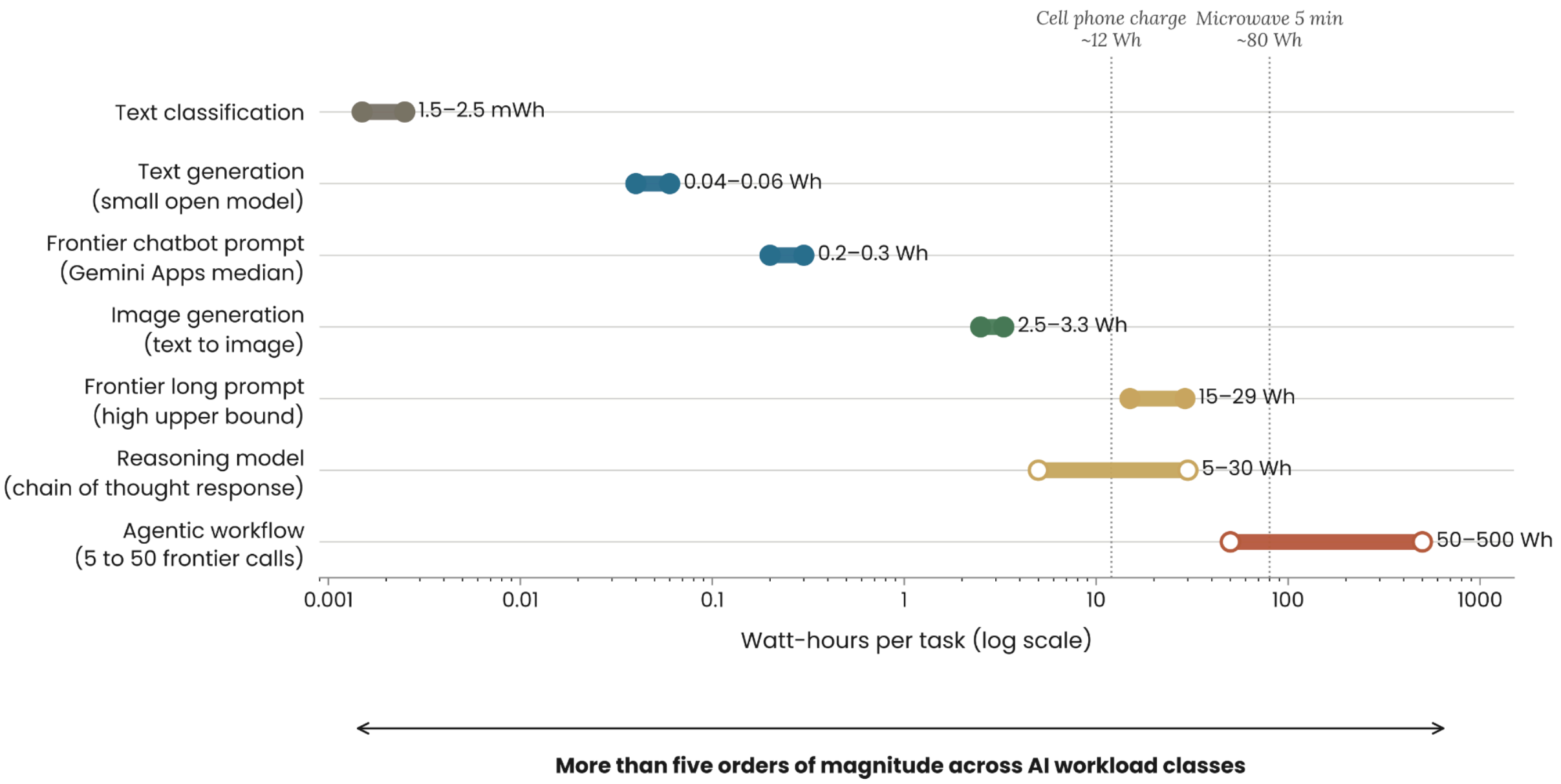

**Figure 2: Electricity consumption per AI task across different workloads.** The closed circles in ranges represent values that come directly from the literature. The open circles represent values derived from per-prompt energy intensities. Data come from Elsworth, et al. (2025) for the frontier chatbot, Jegham, et al. (2025) for reasoning and agentic models, and Luccioni, et al. (2024) for the others.[10]

## 4. Training and Embodied Emissions Are Large But Poorly Allocated

Model training and manufacturing accelerator hardware generate large upfront carbon costs. These emissions must be amortized over time; however, no standard convention exists for how to spread these emissions across use and model lifetimes. Multi-tenant infrastructure further complicates this allocation, as data centers may serve thousands of users and models simultaneously. These issues are compounded by the commercial sensitivity of the denominators involved in the amortization calculation. The total number of tokens a model will serve over its lifetime, duration of a training run, and hardware refresh schedules are all data that providers have business reasons to treat as confidential.

---

[9] These challenges are compounded by providers that do not isolate AI-specific compute from general cloud use.

[10] Luccioni, et al. (2024) values are means across the models benchmarked for each task category. Luccioni, Sasha, et al. "Power Hungry Processing: Watts Driving the Cost of AI Deployment?" *Proceedings of the 2024 ACM Conference* (2024). Frontier long prompt upper bound from Jegham, et al. (2025), "How Hungry Is AI?" Reasoning and agentic model ranges are derived by using the frontier chatbot baseline and Hugging Face AI Energy Score factor. Values reflect the measurement boundaries of the underlying studies, which are generally narrower than the comprehensive boundary proposed in this framework.

For companies with AI use concentrated in widely deployed frontier models, where lifetime token volumes make amortized training relatively small, inference electricity is likely the more material and tractable near-term data gap. Training disclosure remains important for boundary completeness and for models with limited deployment, but the commercial sensitivity of training-related data and maturity of clean energy procurement tools point toward prioritizing inference-side data collection first (see Table 5).

### 5. Electricity Accounting Is Complicated by Clean Resource Procurement and Behind-the-Meter Resources

Many AI providers use power purchase agreements (PPAs) and renewable electricity certificates (RECs). These instruments can reduce market-based emissions even when the local grid is carbon-intensive, which potentially produces a large gap between location- and market-based estimates. It also raises questions about hourly and geographic matching of clean energy to AI workloads as well as dynamic load balancing. Multiple data centers can serve models depending on factors such as the user's location, network latency, and capacity, which makes it exceedingly difficult to trace the grid mix for a given query. Finally, the increased prevalence of behind-the-meter (BTM) generation and energy storage at data centers – power produced on-site without passing through the grid – complicates electricity accounting, which standard factors do not capture. Roughly 30% of planned U.S. data center capacity plans to build BTM generation, with nearly 75% powered by natural gas.[11]

**These challenges mean that estimating GHG emissions from enterprise AI use goes beyond simply applying cloud-accounting methods to a larger workload.** The combination of limited first-party data, unstandardized system boundaries, heterogeneous use cases, opaque provider infrastructure, and contested electricity accounting creates uncertainty. **Each input carries its own error bar, which leads estimates to vary by orders of magnitude.** At the same time, the stakes for getting this right are rising due to the expected growth of AI emissions, which could represent a larger share of corporate emissions footprints within the planning horizon. As that share grows, spend-based approaches become increasingly inadequate, and customers, regulators, and auditors will expect emissions figures that can withstand scrutiny.

---

[11] Because this generation is consumed on-site, it is not reflected in grid-average location-based emissions factors, and its treatment under Scope 2 is unsettled. On-site combustion that the operator owns is conventionally Scope 1, while contractually arranged BTM sits in market-based Scope 2 with no established convention for passing the resulting intensity to enterprise customers. See Cleanview (2025), "Bypassing the Grid: How Data Centers Are Building Their Own Power Plants."

At the same time, **AI investment is becoming a considerable driver of new transmission, firm low-emitting generation, and clean energy procurement**, and a framework that tracks electricity use alongside emissions gives companies and providers data to show that AI demand can contribute to grid decarbonization.

**The methodology proposed here is designed to be defensible within this constraint, tiered to meet companies where their data are, transparent about its assumptions, and built to evolve with greater data availability and model updates.** The framework is also designed for action. Building the methodology around levers that companies can pull is a deliberate choice to keep measurement and mitigation linked as AI capabilities scale.

# Literature Review

## Overview

This section synthesizes peer-reviewed literature, industry publications, and the current state of disclosures. In general, there are two approaches to AI emissions estimates: model-based (which use assumptions about hardware specifications, model parameters, and usage patterns such as token lengths) and empirical measurements (which use standardized hardware and tasks).

Many papers provide incomplete coverage and narrow measurement boundaries (Figure 3). Early papers like Strubell, et al. (2019)[12] focus on model training, including variation across models. Patterson, et al. (2021)[13] illustrate how emissions for model training are 88 times lower than the Strubell, et al. estimate once Google-specific measurements related to data center efficiency, hardware choice, and grid carbon intensity are taken into account. Luccioni, et al. (2022)[14] provide a comprehensive study of the BLOOM model, showing how embodied emissions roughly double the training-only emissions estimate.

---

[12] Strubell, Emma, et al. "Energy and Policy Considerations for Deep Learning in NLP." *Proceedings of the 57th Annual Meeting of the Association for Computational Linguistics* (2019).
[13] Patterson, David, et al. "Carbon Emissions and Large Neural Network Training." arXiv preprint arXiv:2104.10350 (2021).
[14] Luccioni, Sasha, et al. "Estimating the Carbon Footprint of BLOOM, a 176B Parameter Language Model." arXiv preprint arXiv:2211.02001 (2022).

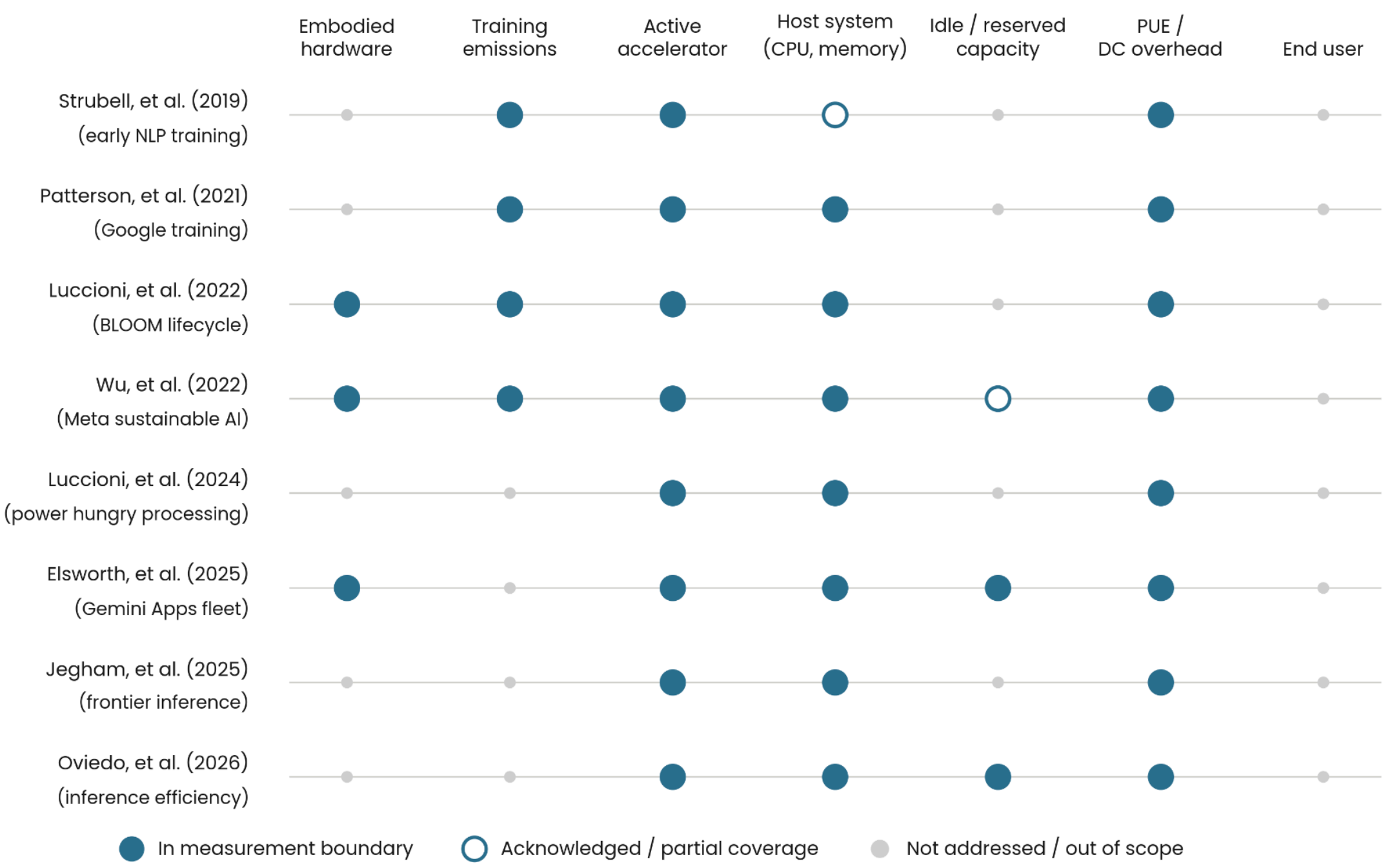

**Figure 3: Boundaries for AI energy and emissions measurements in the literature.** Rows show selected published studies; columns show components of the AI boundary. Filled markers indicate the component sits within the study's measurement boundary; open rings indicate the component is partially addressed but not directly measured. Coverage classifications are based on each study's stated methodology. Adapted from Elsworth, et al. (2025) and extended with subsequent studies.

More recent studies focus on estimates of inference electricity consumption and emissions. Elsworth, et al. (2025) provide a detailed study for Google's infrastructure for serving Gemini, which suggests **order-of-magnitude differences across studies depending on model details, measurement boundaries, and use cases.** Jegham, et al. (2025) combine public performance benchmarks with company-specific environmental multipliers and statistically inferred hardware configurations to estimate inference energy across frontier models, finding that **the most energy-intensive systems exceed 29 Wh per long prompt** and that energy demands scale steeply with parameter count, which highlights the wide variation in emissions intensity that methodologies must accommodate.

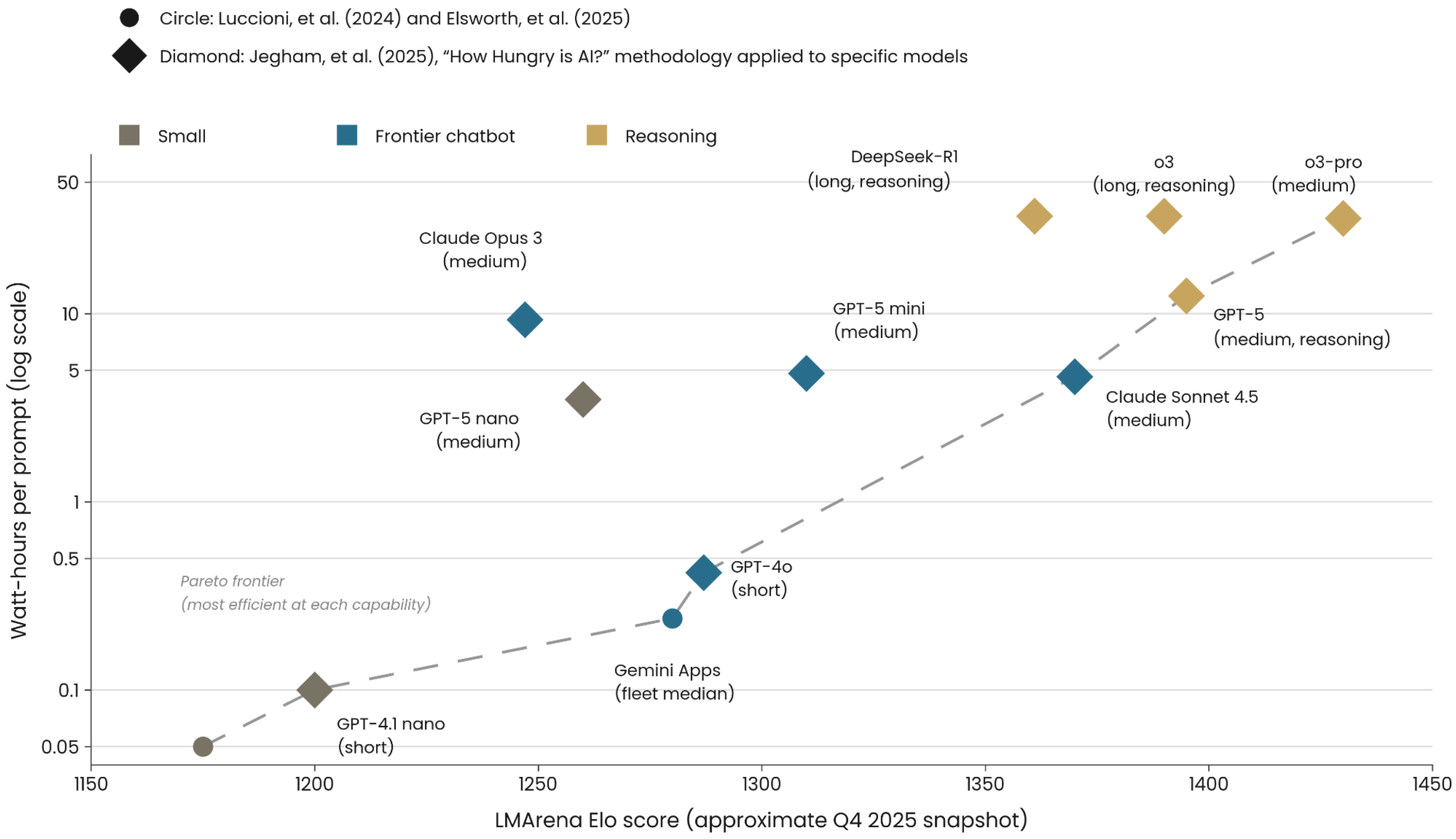

**Figure 4: Electricity consumption per prompt versus capability across large-scale AI models.** Each point represents a model deployed for a text-generation task. Circles denote points from Luccioni, et al. (2024) (direct measurement of open model on standardized hardware) and Elsworth, et al. (2025) (direct disclosure for Google's Gemini Apps serving stack). Diamonds denote points from Jegham, et al. (2025) "How Hungry Is AI?," which estimates per-query inference energy using public API performance benchmarks with statistically inferred hardware configurations and provider-specific multipliers. Color encoding follows Figure 2's workload typology. LMArena scores reflect an approximate Q4 2025 leaderboard snapshot and may shift as new models are added.[15]

Figure 4 plots per-prompt electricity use against capability for select AI models, based on a combination of direct measurements (Luccioni, et al., 2024), provider disclosure (Elsworth, et al., 2025), and estimated values (Jegham, et al., 2025).[16] The figure shows how capability and energy intensity are not strictly coupled, even controlling for prompt complexity.[17] The

[15] Note that the per-query electricity consumption of inference is not a standardized comparison due to the heterogeneity across models, workloads, and measurement approaches.
[16] Although several resources provide model-based estimates of per-prompt electricity consumption, these approximations likely do not reflect efficiencies from frontier models deployed at scale. For instance, Jegham, et al. (2025) estimate roughly 1.6-2.2 Wh for a medium Gemini prompt, an order of magnitude above the 0.24 Wh median reported in Elsworth, et al. (2025).
[17] Fine-tuned small language models (SLMs) and task-specific models can have much lower energy intensity per task than general purpose LLMs for well-defined use cases. This framework focuses on frontier models, because enterprise GHG accounting is largely concentrated in off-the-shelf API consumption from major providers. Companies with material SLM deployments should apply the same tiered methodology using measured or vendor-reported energy intensity figures.

Anthropic time trend (Claude Opus 3 to Sonnet 4.5) shows a two-fold per-prompt efficiency gain alongside ~120 LMArena points of capability improvement over an 18-month period. The GPT-5 family illustrates how scaling with a single model family (nano to mini to full) spans a three- to four-fold energy range. The figure also demonstrates how **reasoning increases per-prompt energy by another order of magnitude.**

Specific energy and emissions intensities ultimately depend on AI usage patterns (e.g., prompt complexity, token length) and AI characteristics (e.g., number of parameters, latency to first-token generation, tokens-per-second, batch size), all of which change frequently by model. Earlier studies did not use models near the Pareto frontier of performance efficiencies, and the use of open models for benchmarking also may alter estimates.[18] Oviedo, et al. (2026) provide direct evidence of this gap, finding that **non-production benchmarks overstate inference energy by 4-20 times relative to production deployments**, primarily because open benchmarks assume batch sizes of one and omit optimizations that hyperscalers apply at scale, which leads to a systematic upward bias.

## Areas of Emerging Consensus

- Inference can dominate lifetime emissions for many models, especially as reasoning and agentic models become more prevalent. Agentic and reasoning workflows (e.g., multi-step model chains, tool use, self-reflection loops) can multiply per-task emissions by orders of magnitude relative to a single-turn query, while variation across input and output modalities (text, image, audio, and video) adds further spread in energy and carbon intensity.
- Geographical factors create extensive variation in carbon intensity due to differences in electricity system generation mixes by region.
- Embodied hardware emissions are smaller than model training and inference operational (electricity) emissions.

## Outstanding Challenges and Areas of Active Disagreement

### Conceptual: Training Emissions Amortization and Attribution

Before discussing how training emissions could be amortized, it is important to establish why these emissions belong inside the boundary. From a Life Cycle Assessment (LCA) perspective, a trained model is a form of capital good: a large, upfront investment of compute that produces a durable asset generating output over a service life.[19] Capital goods are included in the inventory boundary unless proven to be negligible,[20] and the illustrative results below confirm that, for frontier models, this threshold is clearly met. The unsettled questions are how to scope the training workload and how to select the amortization value.

[18] Common benchmarks such as those from AI Energy Score have limitations relative to frontier models deployed at scale. For instance, AI Energy Score assumes a batch size of one for all models, which overestimates footprints. Additionally, AI Energy Score reports GPU plus host CPU energy only and excludes facility overhead, which can increase the figure by 25% to 2x depending on the data center configuration, with frontier hyperscalers at the lower end of the range.

The amortization and attribution of training emissions are important unsettled questions in AI emissions accounting, both in terms of the numerator (which activities count as training) and denominator (how to spread a known emissions quantity over a model's deployed life). A one-time training run can emit thousands of tons of GHG emissions (though disclosures are limited for frontier models).[21] Different approaches produce materially different results, which are particularly fraught when models are deprecated mid-life.[22]

For the numerator, the training-side workload boundary, the final training run is only one component of a model's broader development compute.[23] The broader upstream development includes ablation studies, hyperparameter searches, experiments on smaller-scale versions, synthetic data generation, fine-tuning, training runs that are never deployed, and iterative reward-model training via reinforcement learning from human feedback (RLHF). Epoch AI's analysis of IPO filings finds that final training runs account for a minority of total R&D compute.[24] For the denominator, the total number of tokens a model will serve over its lifetime remains commercially sensitive and unverifiable from outside the provider.

---

[19] This is analogous to the physical hardware whose embodied emissions are included in this framework, which the GHG Protocol considers capital emissions. Standard practice capitalizes these emissions and amortizes across an asset's productive life. Model training has a similar rationale of spreading emissions across the output the model serves over its deployed life, not to attribute them to a single query, customer, or reporting period.
[20] See ISO 14040 and ISO 14044.
[21] Meta's Llama 3.3 70B model card is one of the few public estimates of training emissions. Total location-based GHG emissions were reported as 11,390 $tCO_2e$.
[22] For instance, when GPT-3.5 was superseded by GPT-4, GPT-3.5's training emissions should be amortized over its actual deployed life, however brief, rather than the longer service life originally anticipated. Mid-life deprecation raises the per-token training intensity of the superseded model. Training and embodied hardware emissions likely warrant different treatment, given the different expected service lifetimes of models and physical equipment. AWS applies linear (time-based) depreciation to hardware, while Luccioni, et al. use a six-year life for BLOOM's CNRS infrastructure, though frontier models are reported to refresh hardware on shorter cycles. In either approach, time is a proxy for the more fundamental allocation basis: the share of a hardware unit's functional output (e.g., accelerator-hours, energy throughput, tokens processed) consumed by a given workload. Where functional-output data are unavailable, time-based depreciation is a reasonable fallback, but it can misallocate embodied emissions if utilization varies materially over the asset's life.
[23] As pre-training, mid-training, post-training, and continuous-learning phases become larger shares of total compute, training is shorthand for the larger set of model development workloads.
[24] Jean-Stanislas Denain and Cheryl Wu, "Final Training Runs Account for a Minority of R&D Compute Spending," *Epoch AI*, March 23, 2026.

A separate attribution problem arises from distillation, where a large “teacher” model generates outputs used as training data for a "student" model. This creates what LCA practitioners would recognize as a co-product allocation problem, specifically a form of open-loop recycling, where an output of one product system (the teacher’s inference compute) becomes an input to a different one (the student’s training). The teacher model produces two outputs: a deployed model and a training subsidy that enables the student to be trained with less compute, which raises the issue about whether the student should partially inherit the teacher’s training emissions.

A defensible resolution draws on LCA for recycled inputs. The teacher’s inference compute during distillation is a measurable input to the student's training, which is analogous to electricity or hardware depreciation. However, the teacher’s training emissions remain attributed to the teacher alone. This is similar to how recycled aluminum is handled in manufacturing LCA, where the reprocessing energy for the recycled input is counted, but the original smelter's emissions are not reallocated to the downstream product. A desirable property is that distillation is incentivized as an emissions reduction approach, due to the student's lower training compute.[25]

Given these unresolved questions, an interim approach is to **report training emissions as a separate line item rather than rolling them into a per-token operational figure.** This approach preserves transparency and avoids locking in an amortization or scope convention before the field has reached consensus.[26]

### Conceptual: Accounting for Differences in User Behavior

Variation in how employees and customers use AI introduces further complications that can alter a company's aggregate footprint. Output sequence length appears to be among the most reliable proxies for per-query energy consumption, and the spread across query types is large.[27] Agentic workflows compound this further, because a single request can trigger anywhere from 5-50+ model calls through tool use, self-reflection loops, and multi-step reasoning chains. This dynamic suggests that emissions attributed to one "interaction" may understate the true compute consumed by an order of magnitude or more. Additionally, the distribution of energy per prompt is likely to be skewed, with a small number of complex queries accounting for a disproportionate share of consumption.[28] Characterizing and communicating that distribution in a corporate inventory (e.g., mean, median, or percentile-based range) is itself an open question.

---

[25] ISO 14044 and ISO 14067 contain extensive provisions on allocation for open-loop recycling, which is the basis for the treatment proposed here. Near-term estimates will likely undercount training emissions from frontier models, because the upstream multiplier from failed runs and experiments is omitted without provider disclosure, while efficiency gains from distillation are observable in lower training compute. This bias is important to flag in disclosures.
[26] This recommendation is aimed at model developers disclosing their training emissions (as Meta does for Llama). Customers would include that separate figure in their inventory rather than reconstruct it and, in the absence of provider disclosure, would rely on the Activity Tier default.
[27] The Hugging Face AI Energy Score finds that reasoning models consume roughly 30 times more energy on average than standard inference, with the worst-case queries reaching 113 times the baseline: https://huggingface.co/AIEnergyScore.

### Conceptual: Accounting for Electricity Emissions

Electricity emissions accounting for AI workloads inherits unresolved debates in Scope 2 methods and amplifies them. The central question, whether market-based accounting using energy attribute certificates, RECs, or PPAs should be credited against a company's emissions, or whether location-based accounting is more appropriate, is not specific to AI.[29] But the scale of AI providers' clean energy procurement and the opacity of their serving infrastructure raise the stakes.

The GHG Protocol's Scope 2 dual-reporting guidance nominally applies, requiring location- and market-based figures, but no convention exists for how a model provider passes either figure to an enterprise customer that is auditably linked to the customer's usage. Behind-the-meter generation, often natural-gas-fired generation that has become more common as a bridge to power, adds further complexity, as this energy typically falls outside standard grid-mix reporting.

The choice of emissions factor compounds the problem: most academic studies use average grid intensity because it is the most available option and aligns with GHG Protocol location-based accounting, but some specifications also permit long-run marginal rates, which tells a different story about consequential impacts of AI demand.[30] Long-run marginal rates in particular may be more theoretically appropriate for large and growing loads (i.e., because AI data centers are altering investment decisions), but they are rarely used in practice and methodologically contested. The proposed GHG Protocol Scope 2 revision adds further uncertainty, as it may alter the rules governing EAC crediting.

---

[28] This is a pattern Elsworth, et al. note in the context of Gemini serving, where outlier queries vary "significantly over time."
[29] Location-based emissions use the average grid intensity where consumption occurs, and market-based emissions reflect a company's contractual electricity choices. Market-based emissions factors apply the Scope 2 Quality Criteria hierarchy from GHG Protocol Scope 2 Guidance based on contractual instruments. Energy Attribute Certificates (EACs) including Renewable Energy Certificates (RECs) are tradable instruments that reduce market-based Scope 2 emissions when retired against consumption.
[30] Long-run marginal emissions rates reflect the intensity of new generation built or avoided in response to sustained changes in electricity demand, rather than the average mix of the existing grid (the location-based default) or a supplier's contracted mix (market-based). For AI, this framing maps more directly to the intuition that data center load growth is influencing capacity decisions. See Gagnon, Pieter, et al. "Short-Run Marginal Emission Rates Omit Important Impacts of Electric-Sector Interventions." *Proceedings of the National Academy of Sciences* 119 (2022).

In light of these questions, the most defensible near-term approach is to follow existing GHG Protocol Scope 2 guidance and report both location- and market-based figures.[31] A long-run marginal rate could be reported as an optional figure for companies that wish to signal a more forward-looking view of grid impacts. Although AI emissions sit in Scope 3 for most customers, where the GHGP does not yet formally endorse market-based accounting, the framework recommends dual reporting nonetheless on the rationale that market-based figures capture supplier procurement decisions and create an incentive for upstream reductions.[32]

### Data: Obtaining First-Party Data from AI Model Providers

Data is the binding constraint on accurate accounting, and the pace of change in the underlying systems makes this a moving target. Energy consumption scales directly with accelerator utilization, but inference stacks are evolving rapidly and in ways that cut in both directions, as discussed in the introduction. As capabilities advance, the average query may become more compute-intensive as customers shift more work to AI.[33]

## Proposed Framework for Estimating Emissions from AI Use

A defensible and useable framework needs several elements:

1. Appropriate system boundary for AI energy and emissions measurements
2. Common unit of measure tied to transparent allocation rule
3. Tiered model of measurement that meets companies where data are and reflects data availability

This framework is designed to accept better data as providers publish it and to be updated as models change.

[31] A credible market-based figure also requires that clean electricity claims be allocated across a provider's workloads in proportion to consumption rather than selectively assigning to specific products, which could drive a workload's reported emissions toward zero without procuring additional low-emitting resources. To guard against this, providers reporting per-workload market-based factors should disclose the allocation basis.

[32] The 2015 GHG Protocol Scope 2 Guidance introduced the dual-reporting convention and has not been formally extended to upstream categories, though ongoing GHG Protocol Scope 3 revisions could move in this direction.

[33] Other impacts are ambiguous such as adaptive model routing (i.e., where a provider selects a model based on query complexity), which means that customers may not know which model their request was served by, let alone its energy intensity. Mixture-of-experts architectures, which activate only a subset of parameters, add further heterogeneity.

## Element 1: System Boundary

Without a consistent system boundary, emissions estimates are less comparable. **Boundary choices alone can drive order-of-magnitude variation across published figures** (Figure 3). Narrower approaches that count only active accelerator energy will understate AI's environmental footprint relative to ones that include host system overhead, idle capacity, embodied hardware, or training.

The boundary proposed in this framework is designed to be comprehensive and comparable across providers (Figure 5).[34] Consistent with the GHG Protocol, it covers the full upstream and downstream life-cycle of AI service delivery, from hardware manufacturing through inference, to support comparability across providers, models, and reporting periods.

**Table 1: Proposed system boundary for estimating emissions from AI use.**

| Component | In Scope? | Notes |
|---|---|---|
| Embodied hardware | Yes, amortized | Amortize over useful life; based on Open Compute Project (OCP) data[35] |
| Training emissions | Yes, amortized | Reported as a separate line item rather than rolled into the per-token operational figure; amortization basis disclosed (see Outstanding Challenges) |
| Active AI accelerator power | Yes | Primary driver; measured or modeled based on hardware and utilization |
| Host system (CPU, memory, NIC) | Yes | Typically 10-20% of server power |
| Idle/reserved capacity | Yes | Required for accounting of reserved fleet (Google includes this) |

[34] The framework is consistent with the guiding principles of the GHG Protocol Corporate Accounting and Reporting Standard and with ICT-sector guidance that "*all* data center emissions should be allocated to the services that the data center delivers," which we interpret to encompass Scopes 1, 2, and 3 across the full AI service life-cycle (Figure 5).
[35] Embodied hardware includes raw material extraction, transport, and manufacturing. Although end-of-life impacts are excluded from OCP's boundary, these emissions can be included in embodied hardware or reported separately. End-of-life emissions estimates are currently limited, but including these emissions can help to incentivize recycling and other mitigation measures.

| Data center overhead | Yes | Use provider’s reported power usage effectiveness (PUE), or national average if unavailable |
|---|---|---|
| User-side devices | Out of scope | Attributable to the end user's own Scope 2 |

A complete boundary has two dimensions. The first is the energy boundary, which tracks the energy resources for a given unit of compute (Table 1). The second is the workload boundary, which includes computational steps that are counted as part of the service. An inference request may invoke not only the foundational model call but also safety and moderation classifiers, routing to a serving model, and output guardrails. The question grows more consequential on the training side, where research, pre-training, and distillation workloads must be scoped (as discussed in Outstanding Challenges). The workload boundary is less settled than the energy boundary, but the practical takeaway is that a comprehensive energy boundary drawn only on the headline model call will understate the footprint, especially with safety tooling, routing, and agentic orchestration.

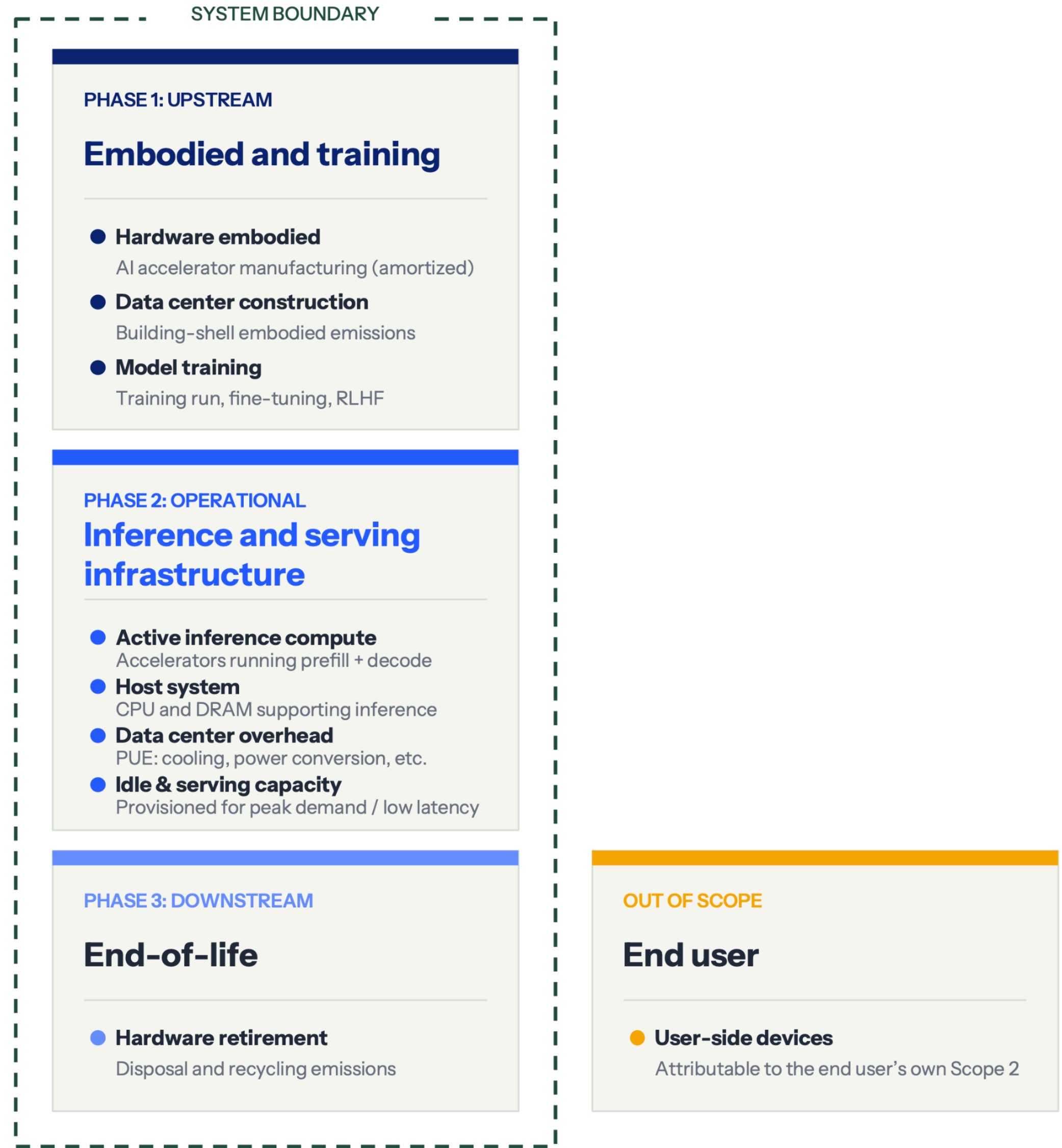

**Figure 5: System boundary for AI emissions framework.**

Embodied emissions from AI accelerator hardware and data center construction are separate boundary decisions.[36] The field is developing norms around how these upfront emissions should be amortized over a device's useful life.[37] Amortization refers to normalizing non-operational emissions outside the use phase across an asset's service life so the full carbon cost is allocated to customers. The Open Compute Project's work on hardware life-cycle accounting provides the most developed framework and is used as the reference point for embodied emissions.[38]

---

[36] Scope 3 embodied carbon is sometimes called "capital emissions" (i.e., similar to financial amortization of capital expenditures), which can include IT hardware, buildings, and other supporting equipment.

[37] AWS assumes linear depreciation, six-year service life for server racks, 50-year service life for buildings, and cradle-to-gate emissions of server racks included. See Amazon (2025), "AWS Customer Carbon Footprint Methodology."

The boundary questions for training-related emissions include not only amortized emissions from the primary training run, but also how to treat failed runs, hyperparameter searches, fine-tuning, and other upstream development activities (discussed further in the Outstanding Challenges section).

On the operational side, the boundary should include inference compute and the full serving stack to deliver that inference to users. A figure that counts only active AI accelerator energy will be lower than one that includes active CPU and DRAM energy, data center overhead captured through PUE (cooling, power conversion, and supporting infrastructure), and idle machine energy. Serving infrastructure is provisioned for peak demand, which means that GPUs run with lower capacity factors to ensure availability and low latency. Google's methodology includes idle capacity, while many other studies do not, which accounts for part of the spread in the literature.[39]

## Element 2: Functional Unit

The choice of functional unit determines how emissions are normalized and compared across providers, models, and use cases. **Kilograms of $CO_2$ equivalent per million tokens ($kgCO_2e$/MTok) is a natural candidate for inference workloads, because tokens are the most common billing and logging metric tracked by major providers and roughly scale with electricity use and emissions.**[40] This makes the unit familiar to enterprise customers and straightforward to populate from existing usage data. Input and output tokens should be tracked and reported separately wherever possible, because they are not energetically equivalent.[41] AI cost management is rapidly maturing as a corporate discipline (e.g., minimizing input length, routing tasks to cheaper models, caching aggressively), and dollars saved through token reduction also reduce emissions. For companies maximizing output per token, the business case for addressing emissions aligns with this operational metric.

---

[38] Open Compute Project (2025), "Embodied Carbon Disclosure Base Specification v1." Vendors generally treat amortization as a downstream issue for customers, which means that the burden of allocating embodied carbon to workloads falls on the method rather than the provider.
[39] This component moves in the opposite direction from the production-efficiency adjustments discussed earlier. For any given figure, the net direction depends on which effects dominate.
[40] One limitation of the token-based unit is that tokenizers vary across models and versions, which means that the same input text leads to different token counts and that token volumes are not fully fungible across providers. In practice, this source of variation is typically smaller than other uncertainties in this framework (e.g., training emissions, grid intensity), and the token remains the most tractable unit for enterprise accounting.
[41] Input tokens are consumed in a parallelized prefill pass, while output tokens are generated sequentially in a more energy-intensive decode pass. A unit that blends the two is a defensible fallback when the split is unavailable, but it will misstate emissions for workloads whose input/output mix departs from the assumed ratio.

The per-token unit can accommodate evolving billing conventions. As providers introduce differentiated pricing for cached input tokens, partial-rerun decode, or similar features, the methodology can apply a provider disclosed factor specific to that token category without changing the framework.[42]

For training and fine-tuning, the framework recognizes that several disclosures can serve this purpose. GPU-hours may provide a more appropriate unit of account for training, reflecting how compute is priced and tracked.[43] The most disaggregated form, exemplified by Meta's per-release Llama disclosures,[44] reports GPU-hours alongside hardware type and resulting $tCO_2e$ on both location- and market-based bases, allowing customers to recompute the figure under alternative grid or amortization assumptions. However, that approach also requires an amortization denominator (e.g., total tokens), which may expose commercially sensitive information about scale and depreciation. Another form embeds training as an amortized per-token contribution, a ratio that plugs into customer calculations without providing the numerator or denominator explicitly. The intent is to lower the bar to first-party disclosure while preserving transparency.

**The framework treats electricity consumption and GHG emissions as two co-equal outputs, both normalized to the per-token functional unit.** Tracking electricity unlocks additional management levers, including clean electricity matching against AI workloads, time- or location-shifted serving toward lower-emitting grids, and tracking of efficiency improvements. Emissions and electricity are linked through the customer's grid factor, and separate reporting can preserve transparency about which lever (efficiency gains or grid decarbonization) is driving annual footprint changes. This choice reflects a fundamental design principle of the framework, namely that **accounting should be built for action.**

[42] A volume-based unit also treats tokens as fungible units of value. However, tokens expended correlate only loosely with task quality, so a volume-only metric could under-penalize verbose, low-value generation. A future unit might normalize by delivered value (e.g., emissions per completed task) rather than volume, but there is currently no auditable measure of output quality.
[43] In contrast, cloud emissions accounting typically normalizes on a set of resource-time units, including vCPU-hour for compute, GB-hour for memory, GB-month for persistent storage residency, and GB transferred for network egress.
[44] Meta's Llama 3.3 model card indicated training emissions of 11,390 $tCO_2e$ (location-based) and 0 $tCO_2e$ (market-based) alongside 7.0M GPU hours.

## Element 3: Multi-Tier Calculation Approach

The framework's data quality hierarchy includes three tiers, which include different levels of input data. Companies should use the highest tier for which reliable data are available, expecting that data availability will improve over time.

### Spend Tier

When data availability is limited, emissions can be approximated by multiplying total AI vendor spend for the reporting period by an economic emissions intensity. AI-related spend is mapped to the closest available BEA sector code and multiplied by an emissions factor from a supply-use input-output model. The current default is 0.134 kg$CO_2$e per 2023 U.S. dollar (CEDA, BEA sector 518200, Data Processing, Hosting, & Related Services).[45] The Spend Tier is the starting point for companies that do not yet have access to activity-level data.

However, this approach carries considerable imprecision. The BEA 518200 emissions factor reflects the average intensity of the data processing and hosting sector, which includes low-margin workloads such as web hosting and managed databases. Such factors do not capture the economics of AI inference, where providers charge a premium per token relative to the energy consumed. This mismatch means **spend-based factors can misestimate true AI emissions by several times in either direction** (depending on the context).[46] A second limitation is data vintage. Underlying input-output tables are often two to five years old, so the factor may not reflect the most recent supply chain structure of data processing activity.

A more tailored factor could be constructed by estimating emissions per dollar of AI service from observed or modeled energy use, emissions intensity, and frontier-model pricing, then translating those into a spend-based coefficient. However, this factor would still have important shortcomings, both in not being company-, model-, region-, or time-specific, and in masking variation in hardware and workload characteristics. Although these limitations make activity-based approaches and provider data preferable, these factors would be improvements over generic data-processing factors, because they would reflect the distinctive workloads and pricing of AI rather than averaging with unrelated digital services.

[45] Open CEDA is offered for free under CC BY-SA license at https://openceda.org/. As an environmentally-extended input-output factor, this value embeds the full upstream supply chain, including embodied emissions from hardware manufacturing and data center construction. Much of this embodied component reflects imported production rather than domestic activity (e.g., overseas semiconductor manufacturing), which implies that the Spend Tier captures embodied emissions in a single factor. Other resources like USEEIO – maintained by the Cornerstone collaboration between Watershed, ERG, and Stanford – have similar emissions factors for the BEA data processing category.
[46] As a complement to the default sector factor, cost-per-activity benchmarks from sources such as the ARC-AGI leaderboard (which reports model-specific inference costs in dollars per million tokens) can be used to cross-check implied energy intensities and assess whether a given provider's pricing is consistent with the sector average assumption embedded in the spend-based factor.

Spend Tier results should be treated as provisional placeholders, given these considerations, and flagged as such in disclosures. These values should be replaced with Activity or Provider Tier data when token-level data becomes available.

## Activity Tier

When provider-reported figures are unavailable, electricity and emissions can be estimated from token volumes using the equations:

```
Electricity (kWh) = (I * EI_in + O * EI_out) * PUE
```

Operational/embodied emissions (kg$CO_2$e) = Electricity * GI / $10^3$ + E_emb * (I + O)

Amortized training emissions (kg$CO_2$e, reported separately) = E_train * (I + O)

where I and O are input and output token volumes,[47] EI_in and EI_out are energy-intensity factors for the prefill and decode phases respectively,[48] PUE is the power usage effectiveness, GI is the grid carbon intensity, and E_emb is amortized embodied emissions. Amortized training emissions are reported as a distinct line item, so that the term (which carries the highest uncertainty) does not obscure the rest. The Customer-Side Telemetry subsection later in the white paper discusses the current state of data availability.

The following default values are proposed for each parameter, presented here for feedback. Final defaults will be published alongside the next version of this framework and refined as the underlying literature evolves.

**Table 2: Default factors for Activity Tier emissions calculations.** Values reflect public data as of June 2026 and should be refreshed as disclosures evolve.

| Parameter | Abbr. | Hyperscaler | Unknown Provider | Source |
| --- | --- | --- | --- | --- |

[47] Where separate input and output token volumes are unavailable, a blended energy-intensity factor can be applied to total volume.
[48] These factors are treated as constant per token, which is a reasonable approximation for dense models. Mixture-of-experts architectures activate only a subset of parameters per token, so factors derived from total parameter counts or dense-model benchmarks may overstate their energy use. A further limitation is that constant per-token factors assume per-token energy is independent of prompt shape. In practice, prefill cost grows super-linearly with input length and decode energy per output token rises with context length, while cached input is cheaper, so a single recombinable pair will tend to undercount long-context and agentic workloads.

| Energy intensity of input tokens (ideally model-specific) | EI_in | 0.32 Wh per 1,000 tokens | 0.70 Wh per 1,000 tokens | Derived from published per-prompt electricity benchmarks |
|---|---|---|---|---|
| Energy intensity of output tokens (ideally model-specific) | EI_out | 0.96 Wh per 1,000 tokens | 2.1 Wh per 1,000 tokens | Consistent with a 3:1 decode-to-prefill electricity ratio[49] |
| Power usage effectiveness | PUE | 1.10 | 1.56 | Hyperscaler default reflects reported averages, while unknown provider reflects global industry average[50] |
| Grid carbon intensity (location) | GI_loc | Regional factor for serving region | 341 $gCO_2e$/kWh (U.S. national average)[51] | Defer to GHG Protocol Scope 2 Guidance source hierarchy; see explanatory note below |
| Grid carbon intensity (market) | GI_mar | Provider-disclosed | 341 $gCO_2e$/kWh (fallback to location-based) | |
| Amortized embodied emissions | E_emb | 0.017 $kgCO_2e$ per million tokens | 0.025 $kgCO_2e$ per million tokens | Based on Open Compute Project methodology[52] |
| Amortized training emissions | E_train | 0.1 $kgCO_2e$ per million tokens | 0.5 $kgCO_2e$ per million tokens | Based on published emissions estimates and lifetime token count[53] |

[49] Patel, Pratyush, et al. "Splitwise: Efficient Generative LLM Inference Using Phase Splitting." *2024 ACM/IEEE 51st Annual International Symposium on Computer Architecture* (2024).
[50] Hyperscaler default reflects 2024-2025 reported averages for Google, AWS, and Microsoft. The unknown provider default reflects the global average from "Uptime Institute Global Data Center Survey 2024."
[51] Ingwersen, Wesley, et al. "Making eGRID 2024 Data Easily Accessible." *Cornerstone*, March 17, 2026.
[52] Values are based on H100-class card at 164 $kgCO_2e$ per NVIDIA Product Carbon Footprint (2025), four-year useful life (consistent with AWS linear depreciation), 50% utilization (midpoint of hyperscaler range from Elsworth, et al., 2025), and representative throughput of 2.4 billion tokens per accelerator-year. Hosts, racks, networking, and data center construction add 25-40% to embodied emissions according to the Open Compute Project methodology. The unknown provider default rate assumes a six-year lifetime (per Luccioni, et al., 2022) and 35% utilization.
[53] Frontier-model training assumed to be 10,000 $tCO_2e$ (based on Llama 3.3 model card) amortized over a central estimate of $10^{14}$ lifetime tokens. The unknown provider value reflects a smaller assumed lifetime token base.

| Unit conversion | N/A | / $10^3$ | | Convert to $kgCO_2e$ |
|---|---|---|---|---|

For the energy intensity of input tokens (prefill), the recommended hyperscaler value of 0.32 Wh per 1,000 tokens is derived from Elsworth, et al. (2025), which reports a median Gemini Apps text prompt of 0.24 Wh, combined with an assumed prompt length of roughly 500 tokens and a 3:1 decode-to-prefill electricity ratio.[54] For the unknown provider, the default of 0.7 Wh/1,000 tokens assumes a less-optimized serving stack (older accelerators, lower batch sizes) and is consistent with the range in Luccioni, et al. (2024).

Electricity consumption (kWh per 1,000 tokens) is the primary Activity Tier output and is calculated from energy intensity and PUE parameters. **Calculations should use model-specific values where possible, given large differences in electricity consumption across models.** The emissions figure is derived by multiplying electricity use by the grid carbon intensity. Both outputs should be reported.

Grid carbon intensity should be sourced according to the GHG Protocol Scope 2 Guidance quality hierarchy, using the most granular factor available for the provider's serving region. Common public sources include eGRID (U.S., subnational) and European Environment Agency (EU member states).

## Provider Tier

The most precise approach is to pull $kgCO_2e$/MTok directly from the AI provider's disclosure or from a machine-readable feed if available. This method is preferred wherever provider data exists, because it embeds the actual energy intensity, grid mix, and embodied hardware allocation for the specific infrastructure serving the customer's workload without the need for external assumptions. Today, Google is the only frontier-model provider to publish per-prompt figures at this granularity (Elsworth, et al., 2025), but its disclosure reports a fleet-wide median for Gemini Apps text prompts that cannot be linked to a specific customer's traffic, model, or region. The convening process described in the later section is designed both to extend equivalent disclosure to other providers, and to surface the company-specific telemetry that customers would need to populate a true Provider Tier estimate.

[54] Where providers expose cache-hit token counts, the prefill term can be disaggregated to reflect the lower energy intensity of cached input, given that it reuses key value states computed in a prior request. This can be important for repeat-context workloads (e.g., agentic loops), where a blended factor would overstate electricity consumption. Elsworth, et al. (2025) do not report a token count for the median prompt; the 500-token figure is a representative assumption, broadly consistent with lengths used in comparable assessments (e.g., Mistral's reported 400-token Le Chat response).

An important consideration for the Provider Tier data sourced through a cloud intermediary (e.g., Gemini Enterprise Agent Platform, Bedrock, Azure OpenAI) is whether the reported figure isolates AI-specific compute from the provider's broader cloud workload, a distinction that hyperscalers handle unevenly and that is still evolving.[55] When provider-reported figures are received as aggregate cloud emissions (i.e., AI and non-AI combined), an allocation step is required to apportion emissions to AI-specific activity. The allocation can be usage-based (cycles, tokens, accelerator-hours) or revenue-based (share of revenue from AI services), and the appropriate choice is often company-specific. Even where AI emissions are isolated, the cloud carbon reports remain monthly rollups rather than per-token figures; combining them with the data described in the Customer-Side Telemetry subsection brings cloud-mediated AI traffic closer to Provider Tier estimates.

The table below summarizes inputs required for each tier. Enterprise-level AI emissions accounting will sum across AI providers, and given how the level of available data may vary across a company's tools, different calculation tiers may be used within an inventory. Starting with the one or two largest sources may capture the bulk of a company's AI footprint and can be expanded over time.[56]

**Table 3: Summary of tier-specific data needs for AI emissions calculations.** Blue cells indicate needed data for given tiers, and orange cells show unneeded data.

| Data input | Spend Tier | Activity Tier (simplified) | Activity Tier (detailed) | Provider Tier |
|---|---|---|---|---|
| Annual API spend ($) | ✓ | ✗ | ✗ | ✗ |
| Total token volume | ✗ | ✓ | ✓ | ✓ |
| Input vs. output token split | ✗ | ✗ | ✓ | ✓ |
| Serving region | ✗ | ✗ | ✓ | ✓ |
| Provider PUE | ✗ | ✗ | Optional | ✓ |
| Grid carbon intensity (location-based) | ✗ | National avg. | Regional | From provider |

[55] AWS's Sustainability Console and Microsoft's Emissions Impact Dashboard provide per-service breakdowns but do not yet distinguish Bedrock or Azure OpenAI as separate lines in their published methodology. Google Cloud's Carbon Footprint methodology applies a single uniform allocation across all covered products but does not define an AI-specific allocation step.
[56] Moving across tiers could trigger a methodology-driven rebaseline. However, rebaselining risk is partially mitigated by the expected growth of AI use, which could dominate a one-time adjustment in many inventories.

| Grid carbon intensity (market-based) | ✗ | ✗ | If provider discloses | ✓ |
|---|---|---|---|---|
| Embodied hardware $CO_2$ | ✗ | Estimated | Estimated | (if included) |
| Amortized training $CO_2$ | ✗ | Approximated | Approximated | (if included) |
| Model family | ✗ | ✗ | ✓ | ✓ |
| Provider g$CO_2$e/token | ✗ | ✗ | ✗ | ✓ |

# Illustrative AI Emissions Calculations

This section applies the framework to illustrate how each tier works, what parameters are required, and how much the estimate changes as more granular data become available.

FinCo LLC is a small financial services company that uses a frontier LLM for internal productivity and as a customer-facing chatbot. The model is accessed via API from a cloud provider's hosted inference infrastructure in the Eastern U.S.

**Table 4: Illustrative data for framework calculations for hypothetical company (FinCo).**

| Parameter | Value | Source |
|---|---|---|
| Annual API spend | $100,000 | Invoice total |
| Input token volume | 13,200 MTok | Based on API usage logs; MTok = millions of tokens |
| Output token volume | 4,400 MTok | API logs |
| Total token volume | 17,600 MTok | Sum of input and output |
| Implied blended price | $2.50/MTok input; $15.00/MTok output | Consistent with frontier-model pricing (approx. Claude Sonnet) |
| Serving region | U.S. East | Cloud provider metadata |
| Model provider | Frontier model API (unnamed) | Exact model and hosting details available only in the Provider Tier |

## Spend Tier (Lowest Precision)

The Spend Tier requires only the dollar amount of AI API spend, which is multiplied by an applicable economic emissions factor. This is an audit-ready starting point for companies that do not yet track AI token use.

| Parameter | Value | Source |
| --- | --- | --- |
| Annual API spend | $100,000 | FinCo invoice |
| Spend-based emissions factor | 0.134 $kgCO_2e$/USD | Open CEDA 2023 value; BEA sector 518200 (Data Processing, Hosting, & Related Services) |

**Spend Tier calculations suggest 13.4 $tCO_2e$** for FinCo's annual AI emissions, which excludes location, usage, and life-cycle data.

This approach carries large uncertainty due to the emissions factor reflecting "data processing and hosting" rather than AI services. However, frontier LLM providers charge a premium per token relative to their energy use, which means that the kWh-per-dollar of AI spend is lower than the sector average. As the tiers below show, this discrepancy causes an overestimate for efficient providers.[57] These effects illustrate why spend-based methods are inadequate for $CO_2$ estimates.

## Activity Tier (Simplified)

Available data on token volumes allow estimates to be grounded in actual compute. The Activity Tier uses a single blended energy-intensity factor (i.e., no distinction between input and output tokens), average PUE, and national average grid carbon intensity. Compared with the Spend Tier, this approach allows emissions to scale with compute consumed rather than spend, which decouples emission estimates from provider pricing and margins.

| Parameter | Value | Source |
| --- | --- | --- |
| Total token volume | 17,600 MTok | FinCo usage logs |
| Blended energy intensity (operational, pre-PUE) | 0.80 Wh / 1,000 tokens | Derived from published range: Elsworth, et al. (2025) Gemini median to Jegham, et al. (2025) high |

[57] Note that, for energy-intensive models on dirty grids, the direction of error can reverse.

| | | |
|---|---|---|
| | | estimates for less-optimized clusters. Use 0.80 Wh as conservative mid-range for frontier model on hyperscaler. |
| PUE | 1.45 | Average U.S. value from LBNL 2024 U.S. Data Center Report |
| Grid carbon intensity (location-based, national) | 341 $gCO_2e/kWh$ | eGRID 2024; location-based without REC credits |

**Simplified Activity Tier calculations suggest 9.1 $tCO_2e$** for FinCo's AI emissions.

This Activity Tier is considerably lower than the Spend Tier, which reflects the fact that spending is at premium prices relative to energy consumed. The estimate is approximate due to its use of blended energy intensity, average PUE, and national-average grid.

## Activity Tier (Detailed)

Additional detail can disaggregate energy intensity and use into separate input-token (prefill) and output-token (decode) factors, use regional grid intensities, and add amortized embodied emissions from hardware and training.

Splitting by input and output token can influence emissions estimates, because input tokens are processed in a parallelized prefill pass that is computationally cheaper per token than the sequential decode (generation) pass. Studies estimate the decode-to-prefill energy ratio at roughly 3:1.

| Parameter | Value | Source |
|---|---|---|
| Input token volume | 13,200 MTok | FinCo usage logs |
| Output token volume | 4,400 MTok | FinCo usage logs |
| Energy intensity input (prefill phase, pre-PUE) | 0.32 Wh / 1,000 tokens | Derived from Elsworth, et al. (2025) and Patterson, et al. (2021) |
| Energy intensity output (decode phase, pre-PUE) | 0.96 Wh / 1,000 tokens | Consistent with Luccioni, et al. (2024) task-level benchmarks and Jegham, et al. (2025) |
| PUE | 1.45 | Average U.S. value from LBNL "2024 United States Data Center Energy Usage Report" |
| Grid carbon intensity (location-based, eGRID RFCE for U.S. East) | 282 $gCO_2e/kWh$ | eGRID 2024 RFCE subregion (representing Virginia data center); lower than national average due to higher nuclear share |
| Grid carbon intensity (market-based) | 141 $gCO_2e/kWh$ | Residual-mix approach; assumes provider holds RECs or PPAs covering 50% of consumption |
| H100 embodied carbon | 164 $kgCO_2e$/card | NVIDIA Product Carbon Footprint disclosure (2025); memory 42% of embodied total |
| Hardware useful life | 4 years | AWS linear depreciation; consistent with Luccioni, et al. (2022) |
| H100 TDP | 700 W per card | NVIDIA H100 SXM5 spec sheet |
| Typical accelerator utilization | 50% | Midpoint of hyperscaler range 40–60%; Elsworth, et al. (2025) |
| Estimated model training emissions | 10,000 $tCO_2e$ | Estimate based on Llama model card location-based estimate |
| Estimated model lifetime token volume | $10^{14}$ tokens | Central estimate: 100 trillion tokens served over model lifetime. Highly uncertain; range of $10^{13}$ to $10^{15}$ explored below. Based on reported usage patterns for major frontier models (e.g., ~100M daily active ChatGPT users, ~500 tokens/query, ~2-year deployment). |

**Detailed Activity Tier calculations suggest 5.4 $tCO_2e$ (location-based) and 3.7 $tCO_2e$ (market-based)** for FinCo's annual AI emissions.

Amortized training adds 1.8 $tCO_2e$, which is roughly 32% of the location-based operational total and about 47% of the market-based one. This fraction rises as the operational footprint shrinks (cleaner grid, more efficient hardware) and falls as the model is used more widely.

**Because the total model token volume is not publicly disclosed by any frontier provider, training carries the highest uncertainty of any term in the calculation.**

### Provider Tier (Highest Precision)

The Provider Tier pulls a published emissions factor directly from the AI provider, bypassing the need to model energy intensity, PUE, or grid mix. This example uses Google's published Gemini Apps figures as a proxy for the Provider Tier.

| Parameter | Value | Source |
|---|---|---|
| Operational energy | 0.24 Wh | Full serving stack: accelerator, host, idle capacity, PUE |
| Carbon emissions (market-based) | 0.03 $gCO_2e$ | Market-based; Google's near-100% renewable PPAs; embodied included |
| Implicit grid intensity | 125 $gCO_2e/kWh$ | Derived: 0.03 $gCO_2e$ / 0.24 Wh * 1,000 |
| Implicit PUE | 1.10 | Google's 2024 global average PUE per Environmental Report |

**The Provider Tier calculations suggest 4.5 $tCO_2e$ (location-based) and 3.2 $tCO_2e$ (market-based)** for FinCo's annual AI emissions, where training is not separately disclosed but embodied emissions are included.

Corporate AI emissions estimates for the illustrative case study are shown in Figure 6. **Training is highly uncertain and contributes 32-56% of the total AI emissions footprint.** The amortized embodied hardware $CO_2$ is much lower than other categories, contributing 4-11% of total AI emissions. Note that, although estimated emissions decline with more granular reporting tiers for this example, this trend may not always hold (e.g., if model training and inference occur in an emissions-intensive grid).

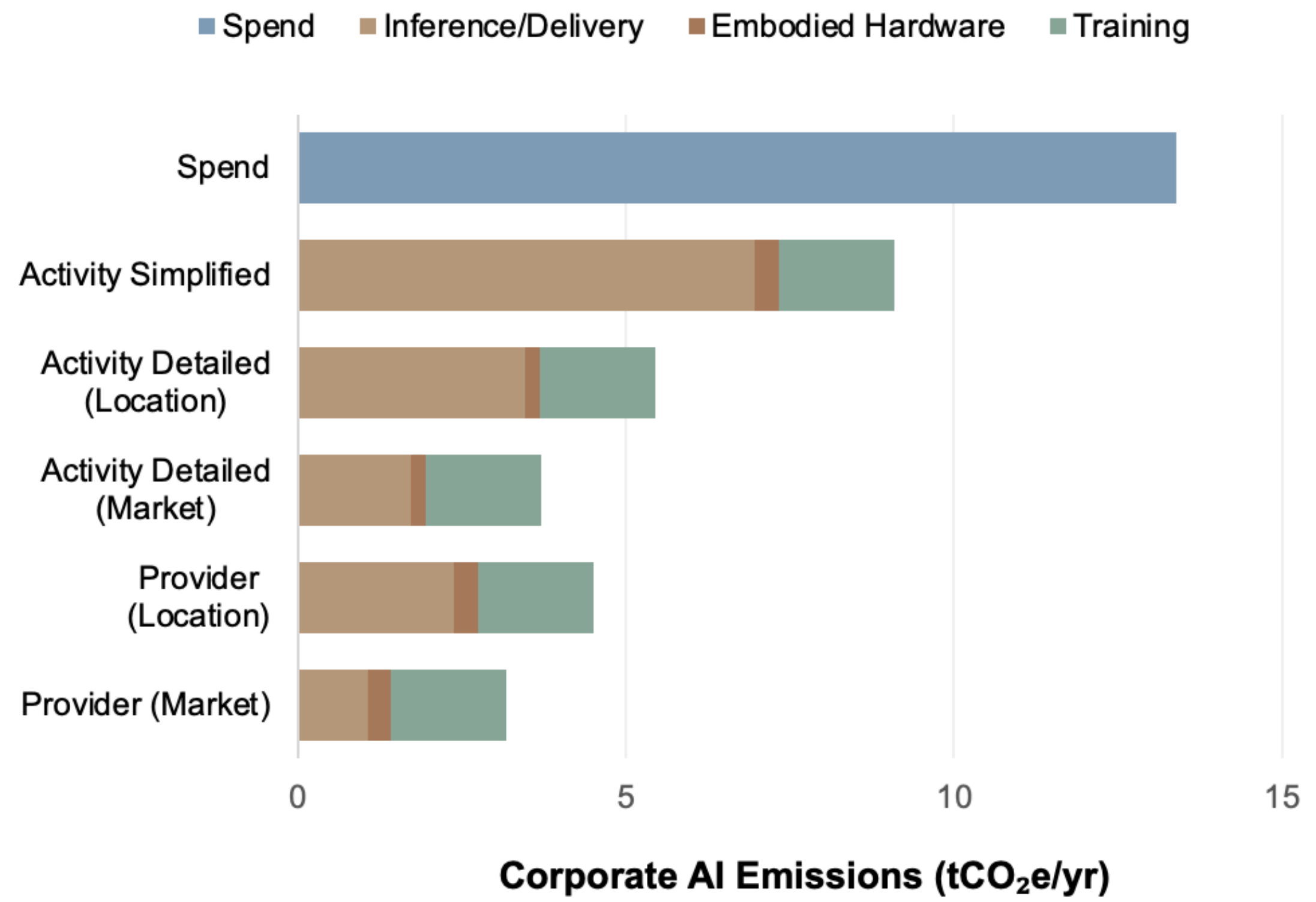

**Figure 6: Summary of emissions estimates for illustrative case study.** See Table 3 for a summary of data sources for each tiered calculation approach.

AI emissions estimates are highly sensitive to uncertain training emissions. The illustrative case study above is based on total emissions estimates from Llama 3.3's model card and central estimate of 100 trillion tokens served. However, neither value is disclosed for frontier models, and as Figure 7 demonstrates, amortizing training emissions over a model lifetime of 10 trillion or 1 quadrillion tokens dramatically changes emissions. Training can be as little as 5% or as much as 83% of total emissions. Although total tokens can be estimated by back-calculating from revenue data, user/usage base, or scaling from open models, each approach leads to wide uncertainty ranges, which means that provider- and model-specific disclosures are important for improving estimates.[58]

[58] Key uncertainties that may put downward pressure on total token estimates include enterprise caching, KV caching within session, and prompt caching. Other uncertainties would increase estimates, including proliferating use cases and agentic workflows for frontier models.

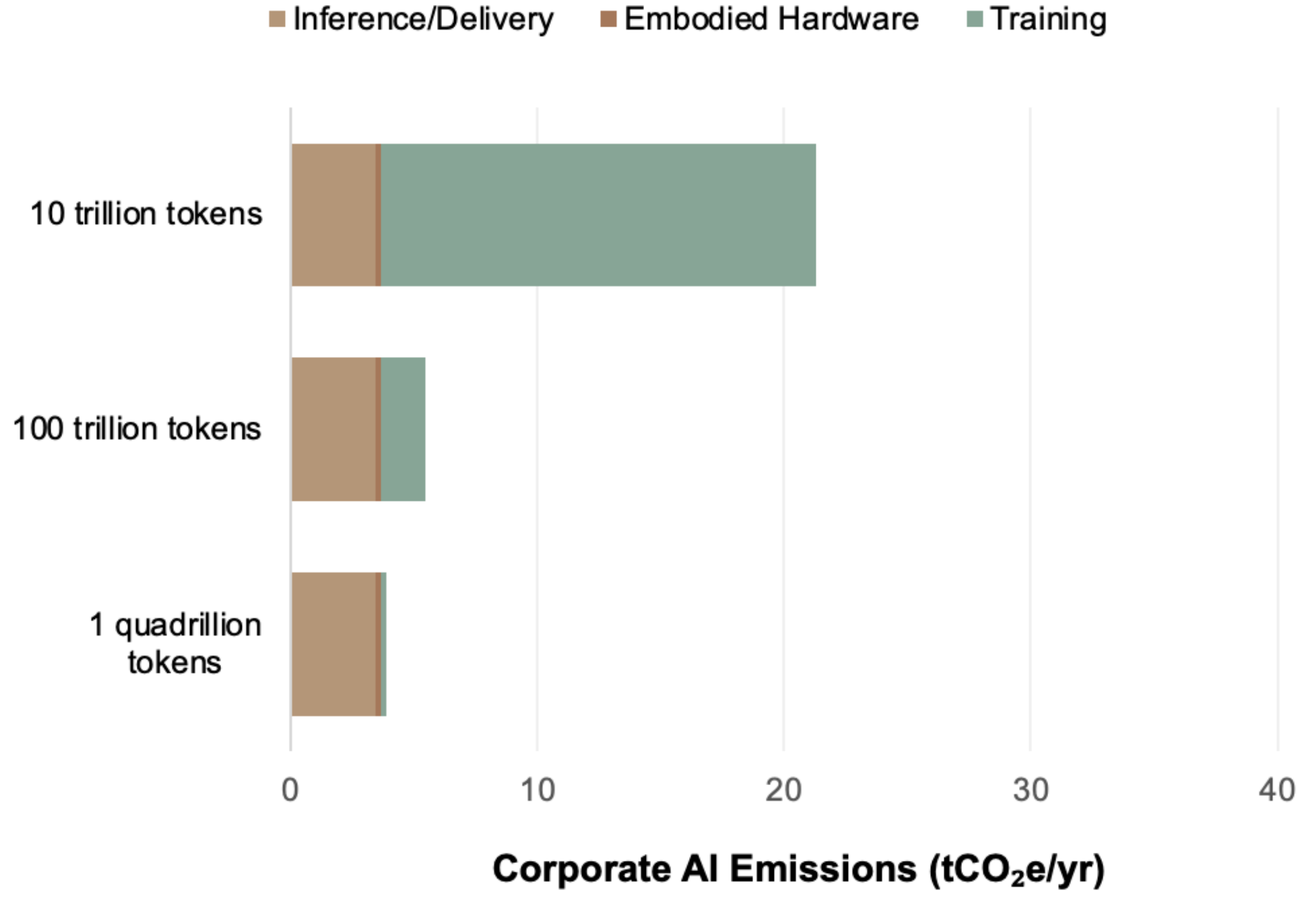

**Figure 7: Emissions estimates under alternate lifetime token count assumptions for illustrative case study.** Values are shown for the detailed Activity Tier calculations with location-based emissions.

# Data Needs and Provider Disclosures

## Customer-Side Telemetry

Inventories need data on both sides, including emissions factors from providers (which is discussed in the next subsection) and customer activity data. However, **activity-based data availability is uneven across channels and providers.**

Channels of corporate AI consumption can be grouped into five categories:

1. **Hosted chat and productivity assistants:** This channel includes vendor-operated chat user interfaces and productivity assistants (e.g., ChatGPT, Claude.ai, Gemini app, Microsoft 365 Copilot). Telemetry is accessible through the vendor's admin console or API, though token-level data are uneven and often hidden behind credits or interaction counts.
2. **Hosted developer tools:** This channel encompasses integrated-development-environment-embedded or command-line interface coding

assistants (e.g., Cursor, Codex, Claude Code, GitHub Copilot). Token-level data are mixed across vendors.

3. **First-party LLM API:** This channel includes customer applications that call the model vendor's API directly with an organization-scoped key. Token, cache, and model data are provided but with limited regional information.
4. **Cloud-mediated LLM API:** This channel encompasses calls through a hyperscaler (e.g., Azure OpenAI in AI Foundry, AWS Bedrock, Vertex AI). Telemetry lives in the hyperscaler's monitoring stack. The serving region is an explicit dimension for this channel, though the model vendor itself has no view into this traffic.
5. **Self-hosted deployments:** Customer-operated inference of downloadable-weights models (e.g., Llama, Mistral). The company holds whatever telemetry it chooses to instrument.

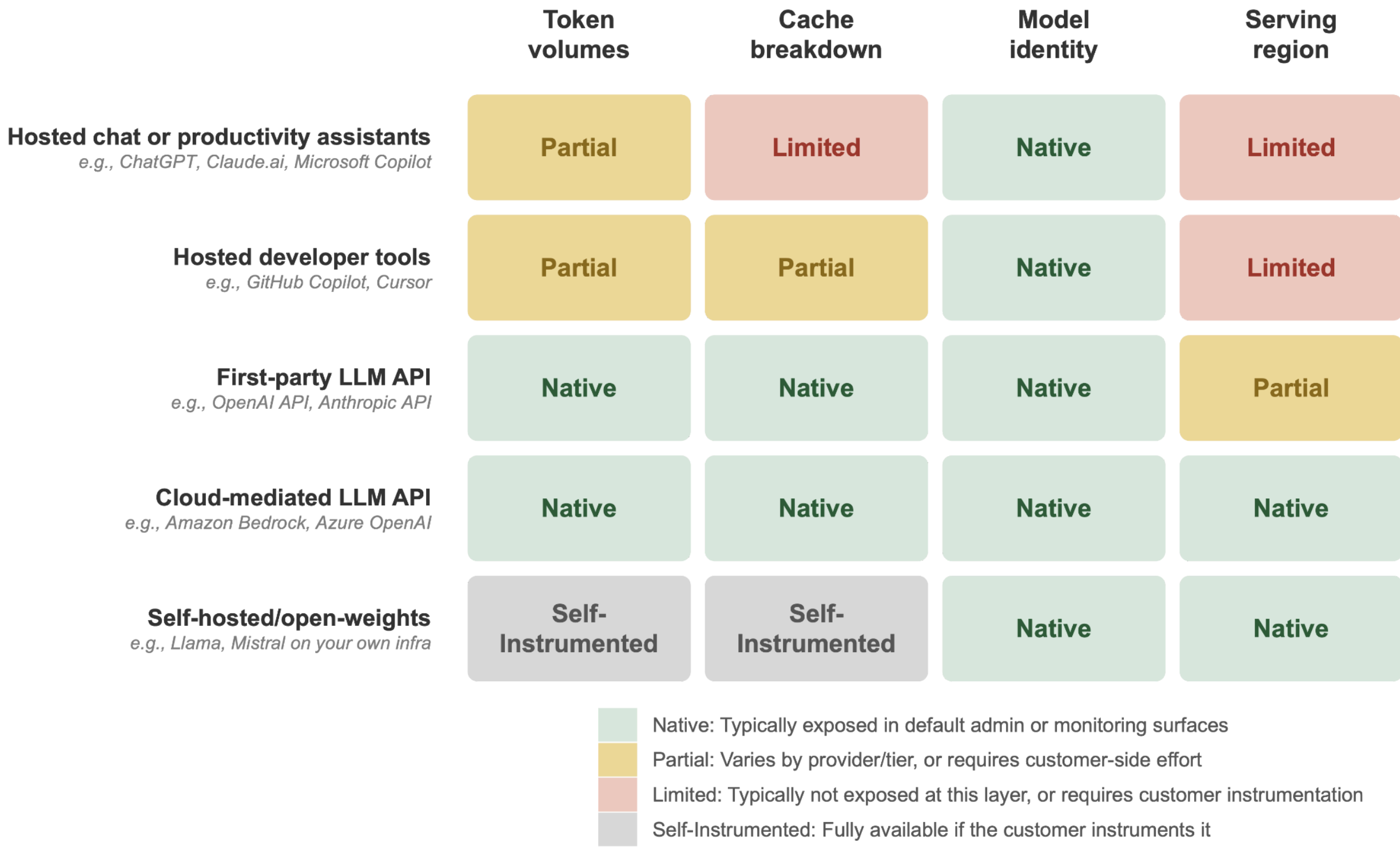

| | Token volumes | Cache breakdown | Model identity | Serving region |
|---|---|---|---|---|
| **Hosted chat or productivity assistants** *e.g., ChatGPT, Claude.ai, Microsoft Copilot* | Partial | Limited | Native | Limited |
| **Hosted developer tools** *e.g., GitHub Copilot, Cursor* | Partial | Partial | Native | Limited |
| **First-party LLM API** *e.g., OpenAI API, Anthropic API* | Native | Native | Native | Partial |
| **Cloud-mediated LLM API** *e.g., Amazon Bedrock, Azure OpenAI* | Native | Native | Native | Native |
| **Self-hosted/open-weights** *e.g., Llama, Mistral on your own infra* | Self-Instrumented | Self-Instrumented | Native | Native |

Native: Typically exposed in default admin or monitoring surfaces
Partial: Varies by provider/tier, or requires customer-side effort
Limited: Typically not exposed at this layer, or requires customer instrumentation
Self-Instrumented: Fully available if the customer instruments it

**Figure 8: Customer-side telemetry and data availability for calculating AI emissions.** Rows show consumption channels (discussed in the list above), and columns represent data dimensions that drive corporate AI emissions calculations.

As shown in Figure 8, customers with material first-party-API or cloud-mediated-API usage can usually populate Activity Tier calculations directly from existing logs. Customers whose AI footprint is concentrated in hosted chat or developer tools often have limited or no visibility into token or credit usage. For these customers, the Spend Tier is the starting point until more granular information is available.[59]

The implication for this framework is that telemetry availability is a per-channel concern as much as a per-provider one. When the same model is accessed through three channels, it can yield three different qualities of input data that can be used to infer emissions.

## Provider Disclosures

The largest constraint on accurate AI emissions accounting is the limited first-party data from companies that operate the infrastructure. Until providers publish these data in standardized and auditable forms, accounting frameworks rely on assumptions that entail order-of-magnitude uncertainty in estimates.

Early academic estimates appear to have overstated AI emissions substantially. Patterson, et al. (2021) find that Google-specific values for data center hardware and grid carbon intensity lower training emissions 88-fold. Elsworth, et al. (2025) estimate a 47-fold reduction in electricity consumption per prompt for Gemini year-over-year. Where AI's true footprint is lower than headlines suggest, **standardized disclosure is the most credible way to say so, replacing assumption-driven estimates with auditable data.** Data transparency therefore serves providers' interest by replacing inflated external estimates, which are attributed to their customers' footprints, with auditable figures that the provider can stand behind.

The Stanford Foundation Model Transparency Index illustrates the extent of the disclosure gap. It finds that the vast majority of frontier models provide no meaningful environmental or energy-related information.[60]

**For many enterprise customers using widely deployed models, inference electricity is likely the higher-priority disclosure** for three reasons. First, for models with larger user bases, lifetime tokens are large enough that amortized training emissions are modest relative to inference (the 32-56% training share in this illustrative example reflects an assumed lifetime token volume at the lower end of the Figure 7 range). Second, inference electricity connects directly with actionable levers such as clean electricity procurement, region selection, and provider efficiency, which are unavailable for sunk training costs. Finally, training-related data is perceived as commercially sensitive information for providers, while per-token inference energy intensity may be more tractable to disclose. Training disclosure

[59] When providers do not charge based on usage, AI activity may be a smaller cost and emissions contributor relative to the supplier's broader footprint, which means the precision gains from moving to the Activity Tier may not justify the effort of reconstructing token volumes from indirect signals.
[60] See https://crfm.stanford.edu/fmti/December-2025/index.html (accessed April 2026).

remains important for boundary completeness and for models with limited deployment, but the field would benefit from prioritizing inference-side data transparency.

## State of Provider Transparency

Existing disclosures are highly uneven across providers with an overall lack of data from frontier models to facilitate more granular customer-level accounting.

Among model developers, Meta is currently the most transparent on training emissions, publishing GPU hours, hardware type, and $tCO_2e$ figures for each Llama release on both a location-based and market-based basis.[61] The contrast between Meta's location-based figure (11,390 $tCO_2e$ for Llama 3.3) and its market-based figure (effectively zero, given renewable electricity procurement) illustrates the gap that dual reporting can reveal.

Google provides the most transparency on the operational side, publishing per-prompt energy and emissions figures for Gemini Apps (Elsworth, et al., 2025). Google Cloud also provides customers with Scope 1, 2, and 3 breakdowns by project, region, and service through its Carbon Footprint tool. However, training emissions are not included.

AWS launched a Sustainability Console in April 2026 providing Scope 1-3 emissions by service and region via API, independently verified by Apex, though without a per-token breakout for inference. Microsoft provides corporate reporting and an Emissions Impact Dashboard for Azure customers but offers no AI-specific breakout of emissions attributable to services such as Azure OpenAI.

## Standards Landscape

More than a dozen working groups claim to be developing AI emissions standards, but none had been widely adopted as of this writing.[62] The Green Software Foundation's SCI for AI specification, ratified in 2025, extends the ISO/IEC 21031:2024 Software Carbon Intensity standard to AI systems. It defines a carbon intensity rate and provides a framework for calculating and comparing emissions across AI deployments. The SCI for AI spec is customer-facing rather than provider-facing, meaning it describes how a customer should structure its accounting rather than what a provider must disclose. The GHG Protocol's ongoing Scope 3 revision (2024 through 2026) is a consequential development for corporate reporting. The Standards Working Group has proposed developing software carbon-intensity specifications that would formally cover cloud and AI services under Category 1 (Purchased Goods and Services). The method proposed here is designed to be consistent with the direction of both efforts.

---

[61] These values exclude non-GPU server consumption, cooling, embodied hardware, and amortized training emissions.

[62] Under the EU Artificial Intelligence Act – Regulation (EU) 2024/1689 – providers of general-purpose AI models must maintain technical documentation that includes a breakdown of the model's energy consumption (Article 53 and Annex XI), which is estimated from computational resources where measured data are unavailable. The obligation applies to GPAI providers generally; energy consumption is also one factor that can contribute to a model's classification as posing "systemic risk," which carries further obligations.

Buyer-side coalitions are also beginning to articulate shared disclosure expectations, signaling demand-side pressure that complements the standards efforts above.[63] This framework is designed to be consistent with that direction.

## What Useful Disclosure Could Include

The fields below describe data that would most improve the accuracy of estimates and, in doing so, give providers a way to replace inflated external estimates with figures they can stand behind. Fields are ordered by priority in Table 5, and not all need to be disclosed simultaneously.[64] Where model providers and their cloud hosts can share these fields (updated on a rolling basis), the resulting estimates become more accurate for the entire industry. The table is offered as a shared reference for what useful disclosure looks like rather than a compliance checklist; providers will be best placed to judge which fields they can share and on what cadence.

**Table 5: Suggested AI model provider disclosures for estimating emissions from AI use.** Rows are ordered by priority: ●●● Essential fields directly enable or materially change customer-level accounting; ●● Important fields improve precision beyond Activity Tier defaults; ● Optional fields support transparency and third-party verification. Providers that can share only a subset of fields should prioritize ●●● rows first, focusing on the fields that deviate most from the Activity Tier defaults in Table 2.

| Field | Priority | Example | Notes |
| --- | --- | --- | --- |
| Model family | ●●● | claude-sonnet-4-6 | Distinct entry per deployed model family |
| Blended energy intensity (Wh per 1,000 tokens) | ●●● | 0.48 | If separate input and output energy intensities cannot be disclosed, a blended intensity can be used as a defensible fallback |

[63] For example, see the Business Council on Climate Change (BC3) AI Sustainability Vendor Engagement Resources, which offer guidance for enterprise buyers engaging AI providers on environmental disclosure: https://www.bc3sfbay.org/aisustainability.
[64] For instance, when a model has lower electricity use per token due to hardware or serving efficiency, providers should make these values transparent so that customer footprints accurately reflect the emissions associated with their specific use rather than a sector-wide approximation.

| | | | |
|---|---|---|---|
| Operational emissions ($kgCO_2e$/MTok) | ●●● | Input: 0.08 and 0.16; output: 0.25 and 0.49 | Market- and location-based (dual-reported per GHG Protocol Scope 2 guidance); reported separately for input and output tokens |
| Region | ●●● | us-east-1 | Physical region of serving, which is important for calculating grid emissions if operational emissions are not given |
| Energy intensity (Wh per 1,000 input tokens) | ●● | 0.40 | Prefill phase; reported separately to enable customer recomputation under alternative grid factors and clean electricity matching |
| Energy intensity (Wh per 1,000 output tokens) | ●● | 1.20 | Decode phase |
| Training emissions ($kgCO_2e$/MTok) | ●● | 0.10 | Amortized training, reported on per-token basis but as a distinct line. Allocation method and assumptions preferred, but given their commercial sensitivity, the per-token ratio is the key quantity for footprints. Reported market-/location-based, because training may occur in different data centers from inference. |
| Behind-the-meter generation (% share) | ●● | 2% gas | Flags on-site generation (e.g., natural gas) serving the workload outside grid average; disclosed separately given distinct treatment and risk exposure, even though emissions are included in the total above |
| Power usage effectiveness (PUE) | ●● | 1.10 | Provider-verified average PUE |
| Reference period | ●● | 2026 Q1 | Time window over which energy intensity and emissions data were measured |
| Valid until | ●● | 2026 Q3 | Data should be refreshed within two quarters of the reference period; aligns with PACT Pathfinder v2 validity period requirements |
| Embodied hardware emissions ($kgCO_2e$/MTok) | ● | 0.017 | Amortized embodied hardware (accelerators plus supporting infrastructure), per-token basis; based on Open Compute Project |
| Methodology version | ● | SCI-for-AI v1.0 | Specification applied (placeholder) |

| Third-party assurance | ● | Apex/DNV | Optional but preferred |
|---|---|---|---|
| Water usage effectiveness (WUE) | ● | 0.12 L/kWh | Average direct (on-site) data center water use, excluding upstream water embedded in electricity generation. Reported separately from $CO_2e$ emissions. Where reported, specify consumption vs. withdrawal. |

# Conclusions and Next Steps

## Practical Approaches to Reducing AI-Related Emissions

AI emissions may currently be a small fraction of footprints for many companies, but converging trends increase expectations for growth. The framework in this white paper aims to establish methods, data, and provider relationships that will be increasingly important over time. The focus is to demystify AI's footprint and embed emissions accounting into AI procurement and engineering decisions.

A practical advantage of the per-token unit is that it aligns with how companies already optimize AI spend. Cost and emissions efficiency share a common metric, which means token-level reporting piggybacks on work companies are already doing. Beyond accounting, the framework supports a small set of practical levers, though they differ in how much effort they require of users and their impacts.

**The highest-leverage lever is decarbonizing data center electricity.** Most of the variation in AI emissions across providers and regions relates to the carbon intensity of electricity consumed. Grid decarbonization and credible clean electricity procurement by data center operators (e.g., co-located renewables and energy storage, hourly matched PPAs, contributions to transmission and firm low-emitting generation) can reduce emissions across all workloads without requiring behavioral changes from users.[65] Although this lever is largely outside of enterprise customers' direct control, it is one the framework is ultimately designed to support: standardized energy intensity disclosure allows customers to select lower-emitting providers and regions, creating market demand from providers that invest in cleaner infrastructure.

[65] EPRI (2026), "Powering Intelligence," EPRI Report 3002034696. Available at: https://powering-intelligence.epri.com/.

The remaining levers reduce electricity consumed per task rather than the carbon intensity of each kWh. They are worth pursuing where they align with existing priorities, but they require added effort to track and implement with potential tradeoffs with productivity and other goals:

- **Optimize the prompt:** The cheapest reductions come before a request is sent, including by writing effective prompts, trimming context windows, eliminating redundant calls, and avoiding agentic loops that do not add value. Output sequence length is among the most reliable proxies for per-query energy, and some use cases may over-generate relative to the user's need.
- **Choose the right model for the task:** Frontier reasoning models can be roughly 30 times more energy-intensive than smaller production models for equivalent queries. Task-aware deployment – routing simple requests to smaller models and reserving frontier capability for when it adds value – is the highest-impact lever available to many customers. However, competitive pressures on capability work against this approach. Unlike most Scope 3 categories (where a primary lever for customers is spend reduction), AI emissions are better addressed through smarter consumption. Spend reduction is a poor proxy, because it could drive customers toward cheaper, less efficient models or self-hosted deployments with worse emissions profiles and no disclosure. Routing potentially aligns customer cost savings, provider utilization, and emissions reduction, making it a rare case where the commercial and environmental incentives point in the same direction.
- **Choose the right region:** In instances where customers can select serving regions, choosing lower-carbon grids reduces emissions. eGRID subregions in the U.S. and EEA country factors in Europe **vary by more than five-fold across available regions.**
- **Choose the right data center in a region:** Where technical capabilities allow, customers can select data centers with credible clean electricity matching and higher efficiencies. This lever requires disclosure that is not yet standard but is part of the framework's requests for providers.

## Broader Context

AI emissions accounting is part of a broader set of public and policy concerns about data center growth. A defensible emissions framework must acknowledge them even where they fall outside the strict GHG inventory.

*Local air quality.* AI infrastructure concentrates load geographically, which can produce criteria pollutant exposure ($NO_x$, $PM_{2.5}$, $SO_2$), particularly where backup generation runs on natural gas or diesel. Carbon-only accounting misses these distributional impacts.

*Water use.* AI workloads can be water-intensive at the cooling stage or entail withdrawals and consumption for electricity generation. The Water Usage Effectiveness (WUE) row included in Table 5 is a starting point that tracks on-site water use. The electricity consumption figures could similarly be used to calculate upstream (off-site) water use associated with power generation; however, this extension is less tractable than the carbon analogue. First, water accounting requires distinguishing withdrawal (i.e., water diverted and largely returned) from consumption (i.e., water evaporated or otherwise removed), which respond differently to cooling technology and carry different local impacts. Second, water intensity varies substantially within a single generation type, which is a function of the cooling system (e.g., once-through, recirculating, dry), ambient climate, and water source, unlike the heat rates and carbon intensities underlying $CO_2$ emissions, which are comparatively uniform across plants of the same fuel. A regional water factor analogous to a grid emissions factor would therefore carry wider uncertainty, and the regionally resolved withdrawal and consumption data needed to construct one are not currently published at the spatial granularity available for location-based grid carbon factors.

*Behind-the-meter generation.* Cleanview (2025) has documented the growth of behind-the-meter generation co-located with new data center capacity, especially using gas-fired resources. It is not yet clear whether these arrangements meet long-term reliability requirements or function as a temporary "bridge to power" until grid connection is possible, and their methodological treatment under GHG Protocol Scope 2 remains contested.

*Grid expansion and modernization.* AI demand is an important potential near-term driver of transmission investment, low-emitting firm generation, and demand flexibility. Beyond its direct electricity demand**, AI could also influence economy-wide emissions through its effects on innovation, technology costs, and energy system efficiency.** Recent modeling using expert elicitations suggests that AI-induced $CO_2$ changes by 2040 range from roughly +11% above to 1.4% below a baseline without AI, depending on the scale of AI deployment and the extent of policy intervention.[66] The wide range reflects uncertainty about whether AI's productivity effects, including potential reductions in the cost of renewables, carbon capture, and nuclear, will outpace its electricity demand. This framework focuses on direct emissions from corporate AI use, but companies and policymakers should weigh this broader context when interpreting per-query or per-company figures in isolation.

*Electricity prices and consumer concerns.* Rising data center demand has become a salient issue in regional rate cases and political debates. Data centers can play constructive roles in these debates through tariff design, transparent reporting, flexibility commitments, and potentially lower electricity prices.[67] Recent research estimates that data centers caused average retail electricity rates to fall modestly in the U.S. from 2015 to 2024 using an instrumental variables approach.[68]

[66] Beath, Hamish, et al. "Artificial Intelligence Drives Divergent Emission Futures" (2026).
[67] See EPRI (2026), "Win-Win Watts: When Can Data Centers, Efficient Electrification, and New Loads Lower Electricity Prices?" EPRI Report 3002034619. Wiser, Ryan, et al. "Factors Influencing Recent Trends in Retail Electricity Prices in the United States." *The Electricity Journal* 38 (2025).

[68] Watten, Asa, et al. "Have Data Centers Raised Your Electric Bill? Causal Evidence from the United States." arXiv preprint arXiv:2606.19777 (2026).

# Authors

Lead author: John Bistline (Watershed)
Contributors: Shaena Ulissi (Watershed), Steven J. Davis (Stanford University), Jonathan Glidden (Watershed), James Joyce (Watershed), Mo Li (Watershed), Jackson Mohsenin (Watershed), Sangwon Suh (Tsinghua University)

# Acknowledgments

We are grateful to the many reviewers across industry, academia, and the standards community who shared feedback on earlier drafts. Their comments improved the framework's rigor, clarity, and practical grounding. Any remaining errors are our own.